\documentclass[reprint,amsmath,amssymb,aps,prb,superscriptaddress,longbibliography,floatfix]{revtex4-2}

\usepackage{mathtools}
\usepackage{graphicx}
\usepackage{dcolumn}
\usepackage{bm}
\usepackage{hyperref}
\usepackage[italicdiff]{physics}

\hypersetup{
	colorlinks=true,
	citecolor=red,
	linkcolor=blue,
	urlcolor=magenta,
}

\begin{document}

\title{Magnetically activated optical visibility of many-body excitons in NiPS$_3$}
\author{Shuta Matsuura}
\affiliation{Department of Physics, University of Tokyo, Hongo, Tokyo 113-0033, Japan}
\author{Shohei Imai}
\affiliation{Department of Physics, University of Tokyo, Hongo, Tokyo 113-0033, Japan}
\author{Naoto Tsuji}
\affiliation{Department of Physics, University of Tokyo, Hongo, Tokyo 113-0033, Japan}
\affiliation{RIKEN Center for Emergent Matter Science (CEMS), Wako, Saitama 351-0198, Japan}
\affiliation{Trans-scale Quantum Science Institute, University of Tokyo, Hongo, Tokyo 113-0033, Japan}

\date{\today}

\begin{abstract}
NiPS$_3$ is a layered van der Waals magnet that provides a unique platform for exploring the interplay between excitons and magnetic order.
In its antiferromagnetic phase, strong many-body interactions give rise to a many-body exciton, which manifests itself as an exceptionally sharp resonance near $1.47 \, \mathrm{eV}$ in optical absorption spectra and photoluminescence.
Previous theoretical studies have assigned the local ground state and the many-body exciton to spin-triplet and spin-singlet states, respectively, both with even parity.
This picture, however, cannot account for the observed optical visibility of the many-body exciton because one-photon transitions between these states are forbidden by spin and parity selection rules.
Here, using group-theoretical analysis and exact diagonalization of single- and two-cluster models, we show that zigzag antiferromagnetic order breaks the relevant symmetries and activates the otherwise forbidden transition. 
We further find that trigonal distortion of the local ligand environment and the inter-cluster exchange interaction relax additional spatial and spin constraints on the transition, producing an exciton signal in the optical conductivity that is distinguishable from the spectral background. 
These results provide a microscopic explanation for the optical visibility of the many-body exciton and its connection to antiferromagnetic order in NiPS$_3$.
\end{abstract}

\maketitle

\section{Introduction}
\label{sec:introduction}
Excitons are electrically neutral excitations that play a central role in the optical response of solids.
In conventional semiconductors, an exciton is typically described as a bound state of an electron and a hole via the Coulomb interaction \cite{wannier1937structure,elliott1957intensity,kazimierczuk2014giant}.
This simple hydrogen picture, however, can break down in strongly correlated materials, where an exciton may become a genuine many-body composite excitation with additional internal degrees of freedom \cite{essler2001excitons,bittner2020photoenhanced,ueda2026doublon,nakamoto2026photoemission}.
In correlated solids, spin--orbit coupling and Hund's coupling \cite{kim2014excitonic,wang2018excitonic,occhialini2024nature}, as well as interactions with collective excitations such as magnons and phonons \cite{gossling2008mott,hariki2020damping,lovinger2020influence,hwangbo2021highly,huang2023spin,mehio2023hubbard,kaneko2023exciton,mehio2025observation}, can play important roles alongside Coulomb interactions, giving rise to excitons with rich internal structures and complex dynamics.

A particularly striking example of such a correlated exciton is found in the layered van der Waals (vdW) antiferromagnet NiPS$_3$, where the Ni ions form a honeycomb lattice and their spins order in a zigzag antiferromagnetic pattern below the N\'{e}el temperature $T_N \simeq 155 \,\mathrm{K}$ \cite{wildes2015magnetic,lancon2018magnetic}.
This correlated exciton manifests itself as a remarkably sharp resonance near $1.47 \,\mathrm{eV}$ in resonant inelastic X-ray scattering (RIXS), photoluminescence (PL), and optical absorption spectra and is commonly referred to as a many-body exciton \cite{kang2020coherent}.
Most notably, the peaks observed in PL and optical absorption disappear above $T_N$, indicating that the optical visibility of the exciton is closely tied to antiferromagnetic order \cite{kang2020coherent}.
Further optical and RIXS studies have shown that the polarization, the magnetic-field dependence, and the dispersion of this excitation are strongly influenced by the antiferromagnetic background \cite{hwangbo2021highly,wang2021spin,kim2023anisotropic,jana2023magnon,song2024manipulation,wang2024unveiling,he2024magnetically}.
These findings have stimulated considerable interest in the microscopic coupling between the many-body exciton and magnetic order.

Previous theoretical studies have investigated the microscopic character of this excitation by calculating the Ni $L$-edge RIXS spectrum within single-cluster NiS$_6$ models \cite{kang2020coherent,he2024magnetically}.
These calculations reproduce key features of the measured spectrum and assign the $1.47\,\mathrm{eV}$ feature to a predominantly local transition from a triplet ground state to a singlet exciton, both of which have even parity.
Such a spin-changing transition can be observed in RIXS because the intermediate state at the Ni $L$ edge contains a $2p$ core hole whose strong spin--orbit coupling permits spin-changing processes \cite{ament2011resonant}.
By contrast, optical absorption and PL do not involve such a strongly spin--orbit-coupled core hole.
Therefore their one-photon transitions usually conserve total spin, and the transition between the triplet ground state and the singlet exciton is expected to be suppressed.
Moreover, because both states have even parity, the parity selection rule forbids a one-photon transition between them.
Consequently, the existing cluster description does not clarify why the exciton produces pronounced peaks in optical absorption and PL.
In addition, because the previous cluster calculations do not incorporate the effect of magnetic order, they do not address why the optical exciton peaks disappear when antiferromagnetic order is lost above $T_N$.

To resolve these issues, we propose that the local symmetry breaking induced by the surrounding zigzag antiferromagnetic order relaxes the relevant spin and parity selection rules and thereby makes the many-body exciton optically active.
To elucidate the essential mechanism in the simplest setting, we first analyze a single-cluster NiS$_6$ model and then extend it to a two-cluster model to develop a more complete description that incorporates inter-cluster exchange coupling. In the single-cluster model, the influence of the surrounding magnetic order is incorporated at the mean-field level.
Group-theoretical analysis and exact diagonalization show that the magnetic mean field produces a finite optical transition matrix element between the low-energy triplet state with $S^{z} = 0$ and the singlet exciton.
Because the transition matrix element vanishes together with the magnetic mean field, this scenario naturally accounts for the disappearance of the optical signal associated with the exciton when antiferromagnetic order is lost.
We further show that trigonal distortion lowers the local symmetry and substantially enhances the otherwise weak optical response.
Although the preceding single-cluster analysis shows that the optical transition from the low-lying triplet state with $S^{z}=0$ to the singlet exciton becomes allowed, the true ground state of this model is the triplet state with $S^{z}=1$, which lies slightly lower in energy than the $S^{z}=0$ state. The optical transition from this true ground state to the singlet exciton is forbidden by conservation of $S^{z}$. 
To overcome this limitation, we consider a two-cluster model and show that quantum spin fluctuations induced by the inter-cluster exchange interaction enable an optical transition from the true ground state to the many-body exciton.

This paper is organized as follows.
In Sec.~\ref{sec:reference_model}, we introduce a reference NiS$_6$ cluster model based on previous studies and show that, in this unperturbed model, the local many-body exciton is optically dark because of spin and parity selection rules.
In Sec.~\ref{sec:single_cluster_oh_activation}, we incorporate the effect of zigzag antiferromagnetic order at the mean-field level and show that the magnetic perturbation relaxes the relevant selection rules.
In the ideal octahedral model, however, the leading nonzero contribution to the optical matrix element is third order in the magnetic perturbation, resulting in a weak optical response.
In Sec.~\ref{sec:single_cluster_d3d_activation}, we show that trigonal distortion lowers the local symmetry from $O_h$ to $D_{3d}$ and allows a contribution linear in the magnetic perturbation, thereby enhancing the optical response.
In Sec.~\ref{sec:two_cluster_model}, we extend the analysis to a two-cluster model and show that exchange-induced quantum spin fluctuations produce a finite optical matrix element between the true ground state and the exciton states.
Finally, Sec.~\ref{sec:summary_and_outlook} summarizes our results and discusses future directions.
Throughout this paper, we set $\hbar = e = 1$.

\section{Reference cluster model and the optically dark exciton}
\label{sec:reference_model}
In this section, we consider a minimal NiS$_{6}$ cluster model for the local many-body exciton in NiPS$_{3}$, based on the cluster models of Refs.~\cite{kang2020coherent,he2024magnetically}. 
Unlike the models used in these studies, the present model omits the Ni $t_{2g}$ orbitals because their hole occupation is small in both the ground and exciton states. We use the ideal $O_h$-symmetric limit without the influence of the surrounding magnetic order as a reference model and show that spin and parity selection rules make the exciton optically dark, in contrast to the exciton peaks observed experimentally in optical absorption and PL.

\subsection{NiS$_6$ cluster model}
\label{sec:reference_model_hamiltonian}
\begin{figure}[tb]
	\centering
	\includegraphics[width=0.98\linewidth]{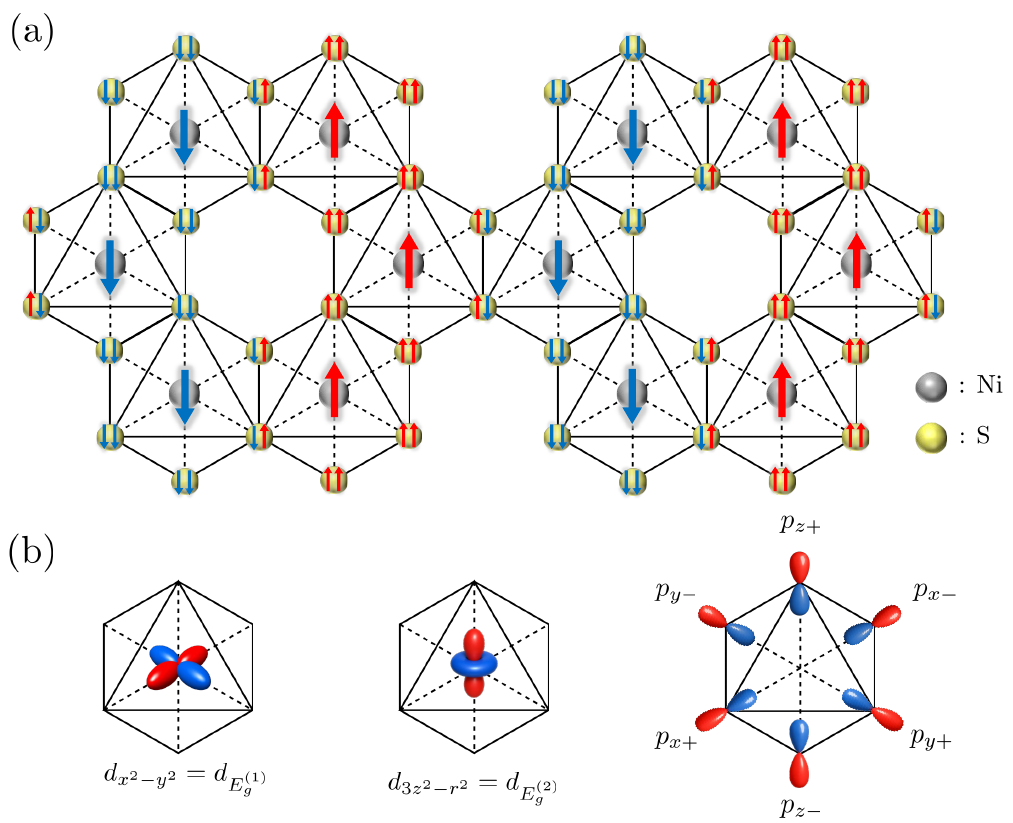}
	\caption{
		(a) Crystal structure of NiPS$_3$.
		The Ni ions form a honeycomb lattice within each layer, and each Ni ion is surrounded by six S ligands in an approximately octahedral NiS$_6$ environment.
		The zigzag antiferromagnetic order below the N\'{e}el temperature $T_N\simeq155\,\mathrm{K}$ is also indicated.
		(b) Orbitals retained in the NiS$_6$ cluster model.
		The model includes the two Ni $e_g$ orbitals and the six ligand $p$ orbitals oriented toward the central Ni ion.
	}
	\label{fig:nips3_structure_and_orbitals}
\end{figure}

NiPS$_3$ is a layered antiferromagnet belonging to the family of transition-metal phosphorus trichalcogenides.
Within each layer, the Ni ions form a honeycomb lattice, and each Ni ion is coordinated by six S ligands, forming an approximately octahedral NiS$_6$ environment, as shown in Fig.~\ref{fig:nips3_structure_and_orbitals}(a).
The actual NiS$_6$ environment deviates from ideal $O_h$ symmetry, with the dominant distortion being trigonal \cite{ouvrard1985structural,kim2019suppression,discala2024elucidating}.
We nevertheless begin with the ideal $O_h$ limit as a reference model, following Ref.~\cite{he2024magnetically}.
This high-symmetry limit provides a convenient starting point for identifying the optical selection rules.
The effect of trigonal distortion is examined separately in Sec.~\ref{sec:single_cluster_d3d_activation}.

In an octahedral crystal field, the five Ni $3d$ orbitals split into the threefold $t_{2g}$ and twofold $e_g$ manifolds.
The nominal Ni$^{2+}$ $3d^8$ configuration is $t_{2g}^{6}e_g^{2}$, with a filled $t_{2g}$ shell and a half-filled $e_g$ shell.
Equivalently, in the hole representation adopted below, this configuration contains two holes in the $e_g$ orbitals.
Hund's coupling favors a local $S=1$ moment, and these local moments develop zigzag antiferromagnetic order below $T_N\simeq155\,\mathrm{K}$ \cite{wildes2015magnetic,lancon2018magnetic}.

To describe the local many-body exciton, we consider a NiS$_6$ cluster consisting of one Ni ion and its six surrounding S ligands \cite{kang2020coherent,he2024magnetically}.
We retain only the two Ni $e_g$ orbitals, $d_{x^2-y^2} \eqqcolon d_{E_g^{(1)}}$ and $d_{3z^2-r^2} \eqqcolon d_{E_g^{(2)}}$, together with one ligand $p$ orbital on each S site oriented along the corresponding Ni--S bond.
The ligand orbitals are denoted by $p_{x\pm}$, $p_{y\pm}$, and $p_{z\pm}$, as illustrated in Fig.~\ref{fig:nips3_structure_and_orbitals}(b).
We work in the hole representation and restrict the Hilbert space to the two-hole sector, corresponding to the nominal Ni$^{2+}$ $d^8$ configuration.
The reference Hamiltonian is
\begin{align}
	H_{0} &= H_{\mathrm{int}} + H_{pd} + H_{pp},
	\label{eq:reference_hamiltonian} \\
	H_{\mathrm{int}}
	&= U_{d} \sum_{\alpha}
	n_{E_{g}^{(\alpha)}\uparrow}
	n_{E_{g}^{(\alpha)}\downarrow}
	+ (U_{d}-2J_{d}) \sum_{\sigma}
	n_{E_{g}^{(1)} \sigma}
	n_{E_{g}^{(2)} \bar{\sigma}}
	\notag \\
	&\quad
	+ (U_{d}-3J_{d}) \sum_{\sigma}
	n_{E_{g}^{(1)} \sigma}
	n_{E_{g}^{(2)} \sigma}
	\notag \\
	&\quad
	- J_{d}
	\qty(
	d_{E_{g}^{(1)} \uparrow}^{\dag}
	d_{E_{g}^{(2)} \downarrow}^{\dag}
	d_{E_{g}^{(2)} \uparrow}
	d_{E_{g}^{(1)} \downarrow}
	+ \mathrm{H.c.}
	)
	\notag \\
	&\quad
	- J_{d}
	\qty(
	d_{E_{g}^{(1)} \uparrow}^{\dag}
	d_{E_{g}^{(1)} \downarrow}^{\dag}
	d_{E_{g}^{(2)} \uparrow}
	d_{E_{g}^{(2)} \downarrow}
	+ \mathrm{H.c.}
	), \\
	H_{pd}
	&= \sum_{\alpha,\mu,\sigma}
	\qty(
	t_{pd}^{\mu\alpha}
	p_{\mu\sigma}^{\dag}
	d_{E_{g}^{(\alpha)}\sigma}
	+ \mathrm{H.c.}
	), \\
	H_{pp}
	&= \varepsilon_p \sum_{\mu,\sigma}
	p_{\mu\sigma}^{\dag}p_{\mu\sigma}
	+ \sum_{\langle\mu,\nu\rangle,\sigma}
	\qty(
	t_{pp}^{\mu\nu}
	p_{\mu\sigma}^{\dag}p_{\nu\sigma}
	+ \mathrm{H.c.}
	).
\end{align}
Here, $\alpha=1,2$ labels the two Ni $e_g$ orbitals, while $\mu=x\pm,y\pm,z\pm$ labels the ligand orbitals.
The operator $d_{E_g^{(\alpha)}\sigma}^{\dag}$ ($d_{E_g^{(\alpha)}\sigma}$) creates (annihilates) a hole with spin $\sigma$ in the Ni $E_g^{(\alpha)}$ orbital, and $p_{\mu\sigma}^{\dag}$ ($p_{\mu\sigma}$) creates (annihilates) a hole in the ligand orbital $p_\mu$.
We use $n_{E_g^{(\alpha)}\sigma}=d_{E_g^{(\alpha)}\sigma}^{\dag}d_{E_g^{(\alpha)}\sigma}$, and $\bar{\sigma}$ denotes the spin opposite to $\sigma$.
The on-site energy of the Ni $e_g$ orbitals is chosen as the energy origin.

The term $H_{\mathrm{int}}$ is the two-orbital Kanamori interaction for the Ni $e_g$ holes \cite{kanamori1963electron}, characterized by the intraorbital Coulomb repulsion $U_d$ and Hund's coupling $J_d$.
The term $H_{pd}$ describes the hybridization between the Ni $e_g$ and ligand $p$ orbitals, with hopping amplitudes $t_{pd}^{\mu\alpha}$.
The ligand Hamiltonian $H_{pp}$ contains the on-site energy $\varepsilon_p$ and the nearest-neighbor hopping amplitudes $t_{pp}^{\mu\nu}$, where $\langle\mu,\nu\rangle$ denotes nearest-neighbor ligand pairs on the NiS$_6$ octahedron.
All nonzero hopping amplitudes are parametrized by the three Slater--Koster integrals $V_{pd\sigma}$, $V_{pp\sigma}$, and $V_{pp\pi}$ \cite{slater1954simplified}.

The Hamiltonian in Eq.~\eqref{eq:reference_hamiltonian} is a minimal version of the cluster models employed in Refs.~\cite{kang2020coherent,he2024magnetically}.
Those models also retain the Ni $t_{2g}$ orbitals.
Ref.~\cite{he2024magnetically}, however, found only a small $t_{2g}$-hole occupation in both the ground state and the local many-body exciton.
The $t_{2g}$ orbitals therefore have little weight in the states relevant to the present analysis, and we omit them from the minimal model.

Because the reference Hamiltonian is invariant under the octahedral group $O_h$, it is useful to introduce symmetry-adapted linear combinations of the ligand orbitals:
\begin{align}
	\begin{pmatrix}
		p_{A_{1g}} \\
		p_{E_g^{(1)}} \\
		p_{E_g^{(2)}} \\
		p_{T_{1u}^{(1)}} \\
		p_{T_{1u}^{(2)}} \\
		p_{T_{1u}^{(3)}}
	\end{pmatrix}
	=
	U
	\begin{pmatrix}
		p_{x+} \\
		p_{x-} \\
		p_{y+} \\
		p_{y-} \\
		p_{z+} \\
		p_{z-}
	\end{pmatrix},
\end{align}
where
\begin{align}
	U =
	\begin{pmatrix}
		\frac{1}{\sqrt{6}} &
		\frac{1}{\sqrt{6}} &
		\frac{1}{\sqrt{6}} &
		\frac{1}{\sqrt{6}} &
		\frac{1}{\sqrt{6}} &
		\frac{1}{\sqrt{6}}
		\\[4pt]
		\frac{1}{2} &
		\frac{1}{2} &
		-\frac{1}{2} &
		-\frac{1}{2} &
		0 &
		0
		\\[4pt]
		-\frac{1}{\sqrt{12}} &
		-\frac{1}{\sqrt{12}} &
		-\frac{1}{\sqrt{12}} &
		-\frac{1}{\sqrt{12}} &
		\frac{2}{\sqrt{12}} &
		\frac{2}{\sqrt{12}}
		\\[4pt]
		\frac{1}{\sqrt{2}} &
		-\frac{1}{\sqrt{2}} &
		0 &
		0 &
		0 &
		0
		\\[4pt]
		0 &
		0 &
		\frac{1}{\sqrt{2}} &
		-\frac{1}{\sqrt{2}} &
		0 &
		0
		\\[4pt]
		0 &
		0 &
		0 &
		0 &
		\frac{1}{\sqrt{2}} &
		-\frac{1}{\sqrt{2}}
	\end{pmatrix}.
\end{align}
The labels $A_{1g}$, $E_g$, and $T_{1u}$ denote the irreducible representations of $O_h$ under which the corresponding ligand combinations transform.
In this symmetry-adapted basis, the hopping terms become
\begin{align}
	H_{pd}
	&= -\sqrt{3}V_{pd\sigma}
	\sum_{\alpha,\sigma}
	\qty(
	p_{E_g^{(\alpha)}\sigma}^{\dag}
	d_{E_g^{(\alpha)}\sigma}
	+\mathrm{H.c.}
	), \\
	H_{pp}
	&= (\varepsilon_p+2T_{pp})
	\sum_{\sigma}
	p_{A_{1g}\sigma}^{\dag}p_{A_{1g}\sigma}
	\notag \\
	&\quad
	+(\varepsilon_p-T_{pp})
	\sum_{\alpha,\sigma}
	p_{E_g^{(\alpha)}\sigma}^{\dag}
	p_{E_g^{(\alpha)}\sigma}
	\notag \\
	&\quad
	+\varepsilon_p
	\sum_{\alpha,\sigma}
	p_{T_{1u}^{(\alpha)}\sigma}^{\dag}
	p_{T_{1u}^{(\alpha)}\sigma},
\end{align}
where $T_{pp}=V_{pp\sigma}-V_{pp\pi}$.
Only the ligand $E_g$ combinations hybridize directly with the Ni $e_g$ orbitals in the reference Hamiltonian.

For the numerical calculations, we adopt the parameter values, $U_d=9.30\,\mathrm{eV}$, $J_d=1.24\,\mathrm{eV}$, $\varepsilon_p=7.5\,\mathrm{eV}$, $V_{pd\sigma}=0.95\,\mathrm{eV}$, and $T_{pp}=1.00\,\mathrm{eV}$, which are close to those in Ref.~\cite{he2024magnetically}.

\subsection{Eigenstates of the Hamiltonian}
\label{sec:reference_model_eigenstates}

We next summarize the local eigenstates of the reference Hamiltonian $H_0$.
Because $H_0$ respects both octahedral $O_h$ symmetry and spin-rotational $\mathrm{SU}(2)$ symmetry, its eigenstates in the two-hole sector can be classified by an irreducible representation of $O_h$, the total spin $S$, and its $z$-component $m$.

The ground-state manifold is a spin triplet with $A_{2g}$ symmetry.
Its three degenerate states are labeled by the $z$ component of the total spin, $m=-1,0,1$, and are denoted by $\ket*{\mathrm{GS},m}$.
Each state can be expressed as a superposition of the $d^8$, $d^9\underline{L}$, and $d^{10}\underline{L}^2$ configuration sectors:
\begin{align}
	\ket*{\mathrm{GS},m}
	={}&
	a_{\mathrm{GS}}
	\ket*{d^8,{}^3A_{2g},m}
	+b_{\mathrm{GS}}
	\ket*{d^9\underline{L},{}^3A_{2g},m}
	\notag\\
	&+
	c_{\mathrm{GS}}
	\ket*{d^{10}\underline{L}^2,{}^3A_{2g},m}.
	\label{eq:ground_state}
\end{align}
Here, $\ket*{d^n\underline{L}^k,{}^{2S+1}\Gamma,m}$ denotes a state with $n$ electrons in the Ni $d$ shell and $k$ ligand holes.
The state transforms according to the irreducible representation $\Gamma$ of $O_h$ and has total spin $S$ with $S^z=m$.
Detailed expressions for the basis states appearing in Eq.~\eqref{eq:ground_state} are given in Appendix~\ref{sec:appendix_eigenstates}.
For the parameters specified in Sec.~\ref{sec:reference_model_hamiltonian}, exact diagonalization gives $a_{\mathrm{GS}}\simeq0.724$, $b_{\mathrm{GS}}\simeq0.670$, and $c_{\mathrm{GS}}\simeq0.163$.
The energy of the ground-state manifold is $E_{\mathrm{GS}}\simeq3.42\,\mathrm{eV}$.

The local many-body exciton is a spin singlet with $A_{1g}$ symmetry.
Its wave function can be expanded in the same charge-configuration sectors as
\begin{align}
	\ket*{\mathrm{EX}}
	={}&
	a_{\mathrm{EX}}
	\ket*{d^8,{}^1A_{1g},0}
	+b_{\mathrm{EX}}
	\ket*{d^9\underline{L},{}^1A_{1g},0}
	\notag\\
	&+
	c_{\mathrm{EX}}
	\ket*{d^{10}\underline{L}^2,{}^1A_{1g},0}.
	\label{eq:exciton_state}
\end{align}
For the same parameters, exact diagonalization gives $a_{\mathrm{EX}}\simeq0.368$, $b_{\mathrm{EX}}\simeq0.894$, and $c_{\mathrm{EX}}\simeq0.256$, with an exciton-state energy of $E_{\mathrm{EX}}\simeq4.87\,\mathrm{eV}$.
The resulting excitation energy is
\begin{align}
	\omega_{\mathrm{EX}}
	=E_{\mathrm{EX}}-E_{\mathrm{GS}}
	\simeq1.45\,\mathrm{eV},
\end{align}
in good agreement with the experimental value $\omega_{\mathrm{EX}} \simeq 1.47\,\mathrm{eV}$ \cite{kang2020coherent}.
The ground state and the exciton therefore have different total spins but share even spatial parity, and these quantum numbers determine the optical selection rules discussed in the next subsection.

\subsection{Optical inactivity of the many-body exciton}
\label{sec:reference_model_selection_rules}

Based on the eigenstates obtained in Sec.~\ref{sec:reference_model_eigenstates}, we now show that, within the reference model, the many-body exciton cannot contribute to the peaks observed in optical absorption or PL.
For light polarized along the $\mu$ direction, the absorption coefficient is related to the longitudinal optical conductivity $\sigma_{\mu\mu}(\omega)$ by
\begin{align}
	\alpha_{\mu}(\omega) \propto
	\frac{\Re \sigma_{\mu\mu}(\omega)}{n_{\mu}(\omega)},
	\label{eq:absorption_coefficient}
\end{align}
where $n_{\mu}(\omega)$ is the corresponding frequency-dependent refractive index.
According to the Kubo formula \cite{kubo1957statistical}, the real part of the optical conductivity is given by
\begin{align}
	\Re \sigma_{\mu\mu}(\omega > 0)
	=
	\frac{\pi}{\omega}
	\sum_{n \neq \mathrm{GS}}
	\abs{\mel{n}{J_{\mu}}{\mathrm{GS}}}^{2}
	\delta(\omega-E_n+E_{\mathrm{GS}}).
	\label{eq:optical_conductivity}
\end{align}
Here $J_{\mu}$ is the $\mu=x,y,z$ component of the electric current operator, which is defined by
\begin{align}
    J_{\mu} = i \qty[H, \, \sum_{a,\sigma} R_{a\mu} n_{a\sigma}],
\end{align}
where $R_{a\mu}$ is the $\mu$ component of the position of orbital $a \in \{ E_{g}^{(1)}, E_{g}^{(2)}, p_{x\pm}, p_{y\pm}, p_{z\pm} \}$.
$\ket{\mathrm{GS}}$ is a ground state, and $\ket{n}$ is the $n$th excited state with energy $E_n$.
The many-body exciton can therefore contribute to optical absorption only if $\mel*{\mathrm{EX}}{J_{\mu}}{\mathrm{GS},m}$ is nonzero for at least one ground state $\ket*{\mathrm{GS},m}$.

We next consider PL, in which the many-body exciton state $\ket*{\mathrm{EX}}$ decays to a ground state $\ket*{\mathrm{GS},m}$ by emitting a photon.
Fermi's golden rule gives the spontaneous emission rate for light polarized along the $\mu$ direction as
\begin{align}
	\gamma_{\mathrm{EX} \to \mathrm{GS},m}^{(\mu)}
	\propto
	\omega_{\mathrm{EX}}
	\abs*{\mel*{\mathrm{EX}}{J_{\mu}}{\mathrm{GS},m}}^{2}.
\end{align}
Thus, the same transition matrix element determines whether the many-body exciton is visible in optical absorption and PL.

In the present model, however, this matrix element vanishes for all $\mu$ and $m$ as a consequence of both spin and parity selection rules.
The first constraint comes from the conservation of total spin.
The reference Hamiltonian is invariant under global spin rotations, and the current operator $J_{\mu}$ is spin independent.
The current operator therefore cannot connect states with different total spins.
Since the ground states have $S=1$, whereas the exciton state $\ket*{\mathrm{EX}}$ has $S=0$, it follows that $\mel*{\mathrm{EX}}{J_{\mu}}{\mathrm{GS},m}$ vanishes for every $m$.

In addition to the spin selection rule, the matrix element is also forbidden by parity.
The current operator $J_{\mu}$ is odd under inversion about the Ni site and can connect only states with opposite parity.
Both the ground states and the exciton state have even parity, as indicated by the $g$ subscripts in $A_{2g}$ and $A_{1g}$.
The matrix element is therefore also forbidden by parity.
Consequently, the reference model cannot account for the many-body exciton peaks observed in optical absorption and PL \cite{kang2020coherent,belvin2021exciton}.

\section{Optical activation of the exciton by magnetic order}
\label{sec:single_cluster_oh_activation}
In this section, we incorporate the influence of the surrounding zigzag antiferromagnetic order at the mean-field level and show that the resulting symmetry breaking activates the otherwise forbidden optical transition to the exciton.
We use group-theoretical analysis to identify how the magnetic perturbation relaxes the relevant selection rules and exact diagonalization to evaluate the resulting transition matrix elements and optical conductivity.

\subsection{Mean-field treatment of magnetic order}
\label{sec:single_cluster_oh_mean_field}

\begin{figure*}[t]
	\centering
	\includegraphics[width=0.85\linewidth]{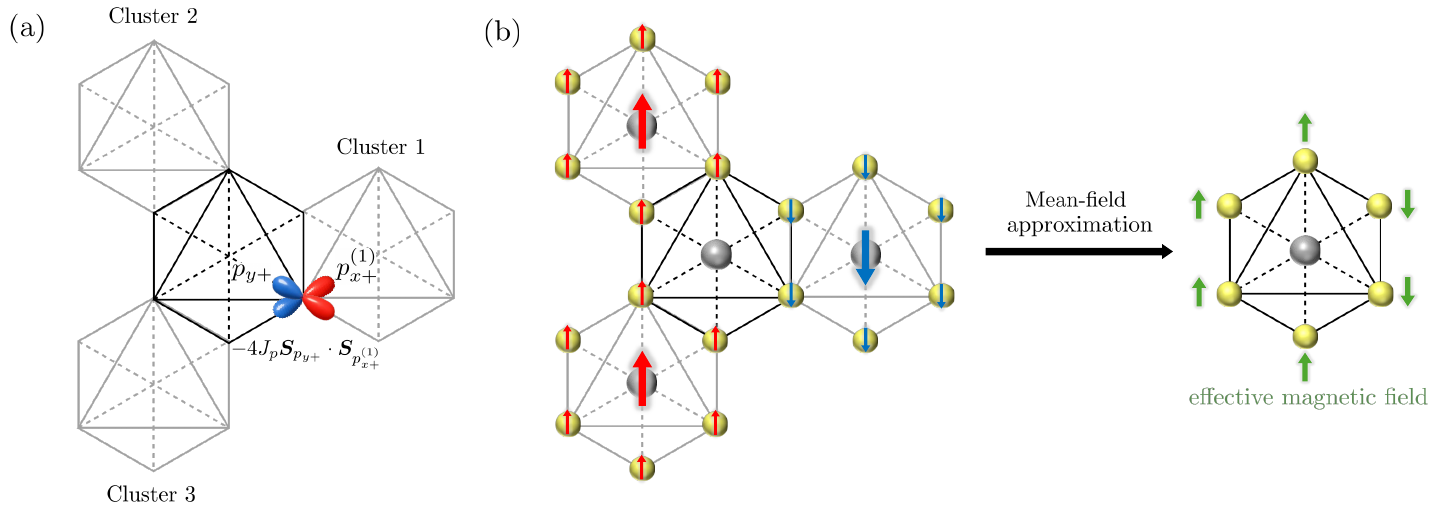}
	\caption{
		(a) Central NiS$_6$ cluster and its three neighboring clusters used in the mean-field treatment of the zigzag antiferromagnetic order.
		At the S ligand shared by the central cluster and cluster 1, the $p_{y+}$ orbital of the central cluster and the $p_{x+}^{(1)}$ orbital of cluster 1 are distinct $p$ orbitals coupled by the Hund interaction.
		(b) Mean-field treatment of the Hund interactions between the ligand $p$ orbitals of the central and neighboring clusters.
		Replacing the spin of each neighboring orbital by its ordered expectation value produces a nonuniform effective magnetic field acting on the ligand $p$ orbitals of the central cluster.
	}
	\label{fig:single_cluster_oh_mean_field}
\end{figure*}

To incorporate the effect of the zigzag antiferromagnetic order, we consider the central NiS$_6$ cluster together with the three neighboring clusters labeled by 1, 2, and 3 in Fig.~\ref{fig:single_cluster_oh_mean_field}(a).
The magnetic moments on the Ni sites in the neighboring clusters are arranged according to the zigzag pattern as shown in Fig.~\ref{fig:single_cluster_oh_mean_field}(b).
Although the ordered moments reside primarily on the Ni $e_g$ orbitals, the $d$-$p$ hopping transfers part of the spin polarization to the ligand $p$ orbitals, inducing a finite ligand spin polarization.
We denote a ligand $p$ orbital in neighboring cluster $i$ by $p_{\nu}^{(i)}$, where the superscript specifies the cluster index, and write its ordered moment as
\begin{align}
	\ev*{S^z_{p_{\nu}^{(i)}}} = \pm m_p.
\end{align}
Here, $m_p>0$ denotes the magnitude of the ligand spin polarization and the sign is determined by the zigzag spin configuration.

In the present model, each S atom belonging to the central cluster is also shared with a neighboring cluster, and the model retains two distinct ligand $p$ orbitals: $p_{\mu}$, which points toward the central Ni ion, and $p_{\nu}^{(i)}$, which points toward the Ni ion in neighboring cluster $i$.
For example, Fig.~\ref{fig:single_cluster_oh_mean_field}(a) shows this configuration at the S site shared by the central cluster and cluster 1, with $p_{y+}$ pointing toward the central Ni ion and $p_{x+}^{(1)}$ pointing toward the Ni ion in cluster 1.
Because these orbitals belong to the same S site, their spins are coupled by the intra-atomic Hund interaction,
\begin{align}
	-4J_p \bm{S}_{p_{y+}} \cdot \bm{S}_{p_{x+}^{(1)}},
\end{align}
where
\begin{align}
	\bm{S}_{p_{\mu}} = \frac{1}{2} \sum_{\alpha,\beta} p_{\mu\alpha}^{\dag} \bm{\sigma}_{\alpha\beta} p_{\mu\beta}
\end{align}
is the spin operator of the $p_{\mu}$ orbital, and $J_p$ is the Hund coupling between ligand $p$ orbitals.
The $p_{x+}^{(1)}$-orbital belongs to the magnetically ordered neighboring cluster, and its spin can be replaced by its ordered moment within the mean-field approximation.
For the pair shown in Fig.~\ref{fig:single_cluster_oh_mean_field}(a), $\ev*{S^z_{p_{x+}^{(1)}}}=-m_p$, and thus
\begin{align}
	-4J_p \bm{S}_{p_{y+}} \cdot \bm{S}_{p_{x+}^{(1)}}
	\simeq 4J_p m_p S_{p_{y+}}^z.
\end{align}

Applying the same mean-field decoupling to all ligand orbital pairs shared with the neighboring clusters yields
\begin{align}
	&H = H_{0} + H_{\mathrm{mag}}, \label{eq:single_cluster_oh_hamiltonian} \\
	&H_{\mathrm{mag}} = h_{\mathrm{eff}} \big(
	S^{z}_{p_{x-}} + S^{z}_{p_{y+}} - S^{z}_{p_{x+}} \notag \\
	&\hspace{2.55cm} - S^{z}_{p_{y-}} - S^{z}_{p_{z+}} - S^{z}_{p_{z-}} \big),
    \label{eq:magnetic_field_pattern}
\end{align}
where $h_{\mathrm{eff}} = 4J_p m_p$.
The nonuniform sign pattern of this effective field is inherited from the zigzag order of the neighboring clusters.

Because the effective field is collinear with the $z$ direction, the Hamiltonian $H$ conserves $S^z$ but is no longer invariant under arbitrary spin rotations.
In addition, the nonuniform sign pattern imposed by the zigzag order is not invariant under inversion around the central Ni site.
Consequently, the spin and parity selection rules that prevent the optical excitation of the exciton in the reference model are relaxed.
Importantly, the relaxation of the parity selection rule relies on the zigzag pattern.
A simple N\'{e}el-type antiferromagnetic order would preserve local inversion symmetry about the central Ni site and therefore leave the parity selection rule intact.

This mechanism is also consistent with the experimental observation that the exciton becomes optically active only below the N\'eel temperature $T_N \simeq 155$ K \cite{kang2020coherent,belvin2021exciton}.
In the present model, the disappearance of the zigzag order above $T_N$ causes $m_p$, and hence $h_{\mathrm{eff}}$, to vanish.
The spin and parity selection rules are then restored, and the one-photon optical transition to the exciton is no longer allowed.

To estimate the magnitude of $h_{\mathrm{eff}}$, we first calculate the ligand spin polarization in the absence of the effective magnetic field.
For the triplet ground state with $S^z=1$, we obtain $m_p \simeq 0.0419$.
Using $J_p=0.3\,\mathrm{eV}$, as estimated from the ligand interaction parameters adopted in Ref.~\cite{he2024magnetically}, the relation $h_{\mathrm{eff}}=4J_pm_p$ yields $h_{\mathrm{eff}}\simeq0.05\,\mathrm{eV}$.
A fully self-consistent calculation would require recalculating $m_p$ in the presence of this effective field, updating $h_{\mathrm{eff}}$, and repeating this procedure until convergence.
However, since the estimated field is small compared with the hopping and interaction energy scales of the cluster Hamiltonian, further iterations are expected to modify its magnitude only slightly.
We therefore use $h_{\mathrm{eff}}=0.05\,\mathrm{eV}$ in the following calculations unless otherwise specified.

Before concluding this subsection, we briefly comment on how our idea relates to two previous studies \cite{kang2020coherent,ergecen2022magnetically}.
Kang \textit{et al.}~\cite{kang2020coherent} focused on the Hund interaction in ligand $p$ orbitals and showed that, in the presence of zigzag magnetic order, the ligand-hole distribution becomes asymmetric about the central Ni site in both the ground and exciton states.
Although they discussed this charge asymmetry as a possible mechanism for magnetoelectric coupling, they did not examine the effect of this Hund interaction on the optical selection rules.
Motivated by the ligand-hole asymmetry found by Kang \textit{et al.}, we interpret the charge redistribution as evidence that the combined effect of zigzag order and the Hund interaction breaks local inversion symmetry, and propose that the resulting loss of inversion symmetry provides a mechanism for lifting the parity restriction on the optical transition.
By treating the Hund interaction at the mean-field level, we explicitly incorporate the resulting local inversion-symmetry breaking into the cluster Hamiltonian and show that the parity constraint is indeed relaxed.

Erge{\c{c}}en \textit{et al.}~\cite{ergecen2022magnetically} examined a closely related effect of zigzag magnetic order on the optical selection rules.
They argued that zigzag magnetic order breaks local inversion symmetry about the Ni site, thereby making an otherwise electric-dipole-forbidden localized $d$--$d$ transition optically accessible.
This mechanism is closely related to ours in that magnetically induced local inversion-symmetry breaking lifts the parity restriction on an optical transition.
Their analysis, however, focused on the parity restriction on a localized $d$--$d$ transition and did not involve the triplet-to-singlet spin change relevant to the many-body exciton.
Relaxation of the spin selection rule is therefore an additional requirement for the optical activation of the many-body exciton considered here.

\subsection{Group-theoretical analysis}
\label{sec:single_cluster_oh_group_theory}
To determine whether $H_{\mathrm{mag}}$ activates the optical transition to the many-body exciton, we analyze the corresponding optical matrix element using perturbation theory and group theory.
Treating $H_{\mathrm{mag}}$ as a perturbation, we expand the ground states and the exciton state in powers of $h_{\mathrm{eff}}$ as
\begin{align}
	\ket*{\widetilde{\mathrm{GS}},m}
	&= \sum_{n \ge 0} h_{\mathrm{eff}}^{n}
	\ket*{\mathrm{GS}^{(n)},m}, \\
	\ket*{\widetilde{\mathrm{EX}}}
	&= \sum_{n \ge 0} h_{\mathrm{eff}}^{n}
	\ket*{\mathrm{EX}^{(n)}}.
\end{align}
Here, $\ket*{\mathrm{GS}^{(0)},m} \coloneqq \ket*{\mathrm{GS},m}$ and $\ket*{\mathrm{EX}^{(0)}} \coloneqq \ket*{\mathrm{EX}}$ are the unperturbed states introduced in Sec.~\ref{sec:reference_model_eigenstates}, while the terms with $n \ge 1$ denote their $n$th-order corrections.
Because the zeroth-order optical matrix element vanishes, the matrix element between the perturbed states can be written as
\begin{align}
	\mel*{\widetilde{\mathrm{EX}}}{J_{\mu}}
	{\widetilde{\mathrm{GS}},m}
	= \sum_{n \ge 1} h_{\mathrm{eff}}^{n} M_m^{(n)},
\end{align}
where
\begin{align}
	M_m^{(n)}
	= \sum_{k=0}^{n}
	\mel*{\mathrm{EX}^{(k)}}{J_{\mu}}
	{\mathrm{GS}^{(n-k)},m}.
	\label{eq:matrix_element_expansion}
\end{align}

The symmetry constraints on $M_m^{(n)}$ can be obtained by examining the individual terms in Eq.~\eqref{eq:matrix_element_expansion}.
For a given $k$, a necessary condition for
\begin{align}
	\mel*{\mathrm{EX}^{(k)}}{J_{\mu}}
	{\mathrm{GS}^{(n-k)},m}
\end{align}
to be nonzero is that the product
\begin{align}
	\Gamma_{\ket*{\mathrm{EX}^{(k)}}}
	\otimes \Gamma_{J_{\mu}}
	\otimes \Gamma_{\ket*{\mathrm{GS}^{(n-k)},m}}
	\label{eq:corrected_state_product}
\end{align}
contain a nonzero component that transforms as the trivial representation.
Here, $\Gamma_{\ket*{\mathrm{EX}^{(k)}}}$, $\Gamma_{J_{\mu}}$, and $\Gamma_{\ket*{\mathrm{GS}^{(n-k)},m}}$ denote the representations of the $k$th-order correction to the exciton state, the current operator, and the $(n-k)$th-order correction to the ground state, respectively.
According to perturbation theory, the states $\ket*{\mathrm{EX}^{(k)}}$ and $\ket*{\mathrm{GS}^{(n-k)},m}$ are generated by applying $H_{\mathrm{mag}}$ $k$ and $n-k$ times to the unperturbed states $\ket*{\mathrm{EX}}$ and $\ket*{\mathrm{GS},m}$, respectively.
Thus, the symmetry analysis of the product in Eq.~\eqref{eq:corrected_state_product} reduces to examining
\begin{align}
	\Gamma_{H_{\mathrm{mag}}}^{\otimes n}
	\otimes \Gamma_{\ket*{\mathrm{EX}}}
	\otimes \Gamma_{J_{\mu}}
	\otimes \Gamma_{\ket*{\mathrm{GS},m}}.
	\label{eq:perturbative_symmetry_product}
\end{align}
For $M_m^{(n)}$ not to be forbidden by symmetry, this product must contain a nonzero component that transforms as the trivial representation.

The unperturbed Hamiltonian $H_0$ has $O_h$ point-group symmetry and spin-rotational $\mathrm{SU}(2)$ symmetry.
The corresponding transformation properties of the states and operators appearing in Eq.~\eqref{eq:perturbative_symmetry_product} are summarized in Table~\ref{tab:single_cluster_oh_representations}.
Importantly, $H_{\mathrm{mag}}$ contains the odd-parity $T_{1u}$ component under $O_h$ and transforms as the $q=0$ component of a rank-1 spherical tensor under spin rotations.
These transformation properties show that $H_{\mathrm{mag}}$ breaks inversion and spin-rotational symmetries, thereby relaxing the corresponding parity and spin selection rules.
\begin{table}[htbp]
	\centering
	\caption{
		Transformation properties of the states and operators in Eq.~\eqref{eq:perturbative_symmetry_product} under $O_h$ operations and spin rotations.
	}
	\begin{ruledtabular}
	\begin{tabular}{ccc}
		& $O_h$ symmetry & $\mathrm{SU}(2)$ symmetry \\
		\colrule
		$\ket{\mathrm{GS},m}$
		& $A_{2g}$ & $S=1$, $S^z=m$ \\
		$\ket{\mathrm{EX}}$
		& $A_{1g}$ & $S=0$, $S^z=0$ \\
		$J_\mu$
		& $T_{1u}$ & rank $0$, $q=0$ \\
		$H_{\mathrm{mag}}$
		& $A_{1g} \oplus E_g \oplus T_{1u}$
		& rank $1$, $q=0$
	\end{tabular}
	\end{ruledtabular}
	\label{tab:single_cluster_oh_representations}
\end{table}

Evaluating the product in Eq.~\eqref{eq:perturbative_symmetry_product} for $m=0$ using the transformation properties listed in Table~\ref{tab:single_cluster_oh_representations} yields the selection rules for $M_0^{(n)}$ summarized in Table~\ref{tab:single_cluster_oh_selection_rules}.
Only the result for the $m=0$ sector is shown because conservation of $S^z$ forces $M_m^{(n)}$ to vanish for $m \neq 0$ at every order.
At first order, the spin selection rule allows $M_0^{(1)}$, whereas the $O_h$ selection rule prohibits it.
The roles of the two constraints are reversed at second order.
At third order, neither selection rule prohibits the matrix element.
Thus, the leading symmetry-allowed contribution appears at third order, and the matrix element is expected to scale as $h_{\mathrm{eff}}^{3}$ for small $h_{\mathrm{eff}}$.
The detailed derivation of these selection rules is given in Appendix~\ref{sec:appendix_group_theory}.

\begin{table}[htbp]
	\centering
	\caption{
		Point-group and spin selection rules for $M_0^{(n)}$ in the $O_h$ model.
		Here, ``allowed'' means that the contribution is not forbidden by the corresponding symmetry.
	}
	\begin{ruledtabular}
	\begin{tabular}{cccc}
		& $O_h$ symmetry & $\mathrm{SU}(2)$ symmetry & Combined result \\
		\colrule
		$M_0^{(1)}$ & prohibited & allowed
		& $M_0^{(1)}=0$ \\
		$M_0^{(2)}$ & allowed & prohibited
		& $M_0^{(2)}=0$ \\
		$M_0^{(3)}$ & allowed & allowed
		& $M_0^{(3)}$ can be nonzero
	\end{tabular}
	\end{ruledtabular}
	\label{tab:single_cluster_oh_selection_rules}
\end{table}

\subsection{Numerical results}
\label{sec:single_cluster_oh_numerical_results}
\begin{figure}[tb]
	\centering
	\includegraphics[width=0.9\linewidth]{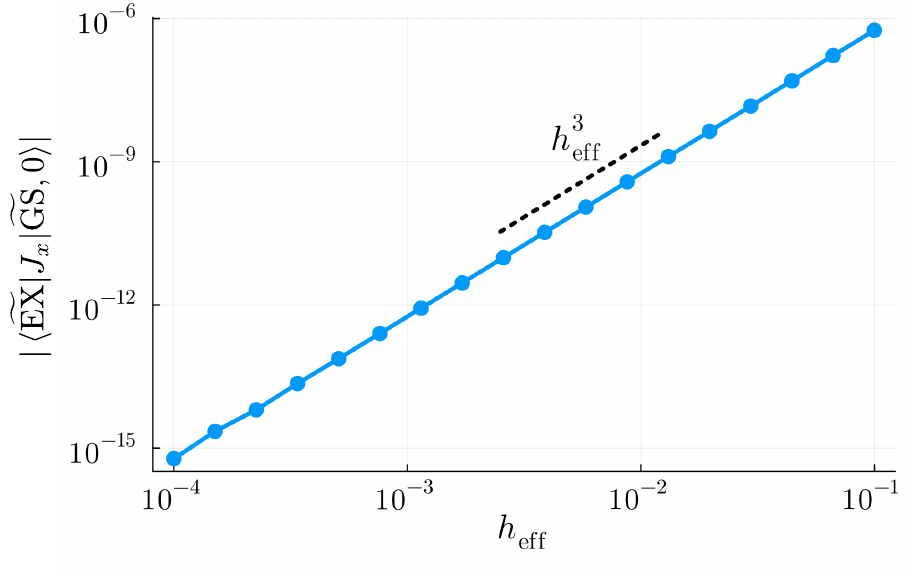}
	\caption{
		Absolute value of the matrix element $\mel*{\widetilde{\mathrm{EX}}}{J_x}{\widetilde{\mathrm{GS}},0}$ for the single-cluster NiS$_6$ model in Eq.~\eqref{eq:single_cluster_oh_hamiltonian} as a function of the effective magnetic field $h_{\mathrm{eff}}$.
		The matrix element scales as $h_{\mathrm{eff}}^3$ in the small-field regime, consistent with the group-theoretical analysis.
	}
	\label{fig:single_cluster_oh_matrix_element}
\end{figure}

The group-theoretical analysis shows that the first- and second-order contributions to the optical matrix element vanish by symmetry, whereas a third-order contribution is not forbidden.
This symmetry condition alone, however, does not guarantee that the third-order coefficient $M_0^{(3)}$ is nonzero.
To determine whether the magnetic perturbation actually activates the exciton, we perform exact diagonalization of the Hamiltonian \eqref{eq:single_cluster_oh_hamiltonian} and directly evaluate the matrix element.
Figure~\ref{fig:single_cluster_oh_matrix_element} shows $\abs*{\mel*{\widetilde{\mathrm{EX}}}{J_x}{\widetilde{\mathrm{GS}},0}}$ as a function of $h_{\mathrm{eff}}$.
The matrix element becomes finite for nonzero $h_{\mathrm{eff}}$ and scales as $h_{\mathrm{eff}}^3$ in the small-field regime.
This result demonstrates that the effective magnetic field activates the optical response of the exciton and confirms that the symmetry-allowed third-order contribution is indeed nonzero.

In addition to the matrix element, we calculate the real part of the optical conductivity $\Re\sigma_{\mu\mu}(\omega)$.
As shown in Eq.~\eqref{eq:absorption_coefficient}, the absorption coefficient $\alpha_{\mu}(\omega)$ and $\Re\sigma_{\mu\mu}(\omega)$ have peaks at the same energies, so we use the optical-conductivity spectrum to discuss the features expected in optical absorption.
The perturbation $H_{\mathrm{mag}}$ lifts the threefold degeneracy of the ground-state manifold of $H_0$, and the true ground state for $h_{\mathrm{eff}}>0$ is given by $\ket*{\widetilde{\mathrm{GS}},1}$.
However, because the Hamiltonian \eqref{eq:single_cluster_oh_hamiltonian} conserves $S^z$, the matrix element $\mel*{\widetilde{\mathrm{EX}}}{J_{\mu}}{\widetilde{\mathrm{GS}},1}$ vanishes, and hence the optical conductivity calculated from the true ground state contains no exciton peak.
For simplicity, we instead regard the low-energy state $\ket*{\widetilde{\mathrm{GS}},0}$ as the ground state in the present single-cluster calculation and evaluate the optical conductivity as
	\begin{align}
		&\Re \sigma_{\mu\mu}(\omega > 0) \notag \\
		&\hspace{0.2cm} = \sum_{n \neq \widetilde{\mathrm{GS}}, 0} \frac{\abs*{\mel*{n}{J_{\mu}}{\widetilde{\mathrm{GS}}, 0}}^{2}}{E_{n} - E_{\widetilde{\mathrm{GS}}, 0}} \frac{\eta}{(\omega - E_{n} + E_{\widetilde{\mathrm{GS}}, 0})^{2} + \eta^{2}},
	\end{align}
where $\eta=10^{-3}\,\mathrm{eV}$ is a broadening factor.
The use of $\ket*{\widetilde{\mathrm{GS}},0}$ instead of the true ground state $\ket*{\widetilde{\mathrm{GS}},1}$ is a limitation of the present single-cluster treatment, and this issue is resolved in Sec.~\ref{sec:two_cluster_model}, where we analyze a two-cluster model.

\begin{figure}[tb]
	\centering
	\includegraphics[width=0.9\linewidth]{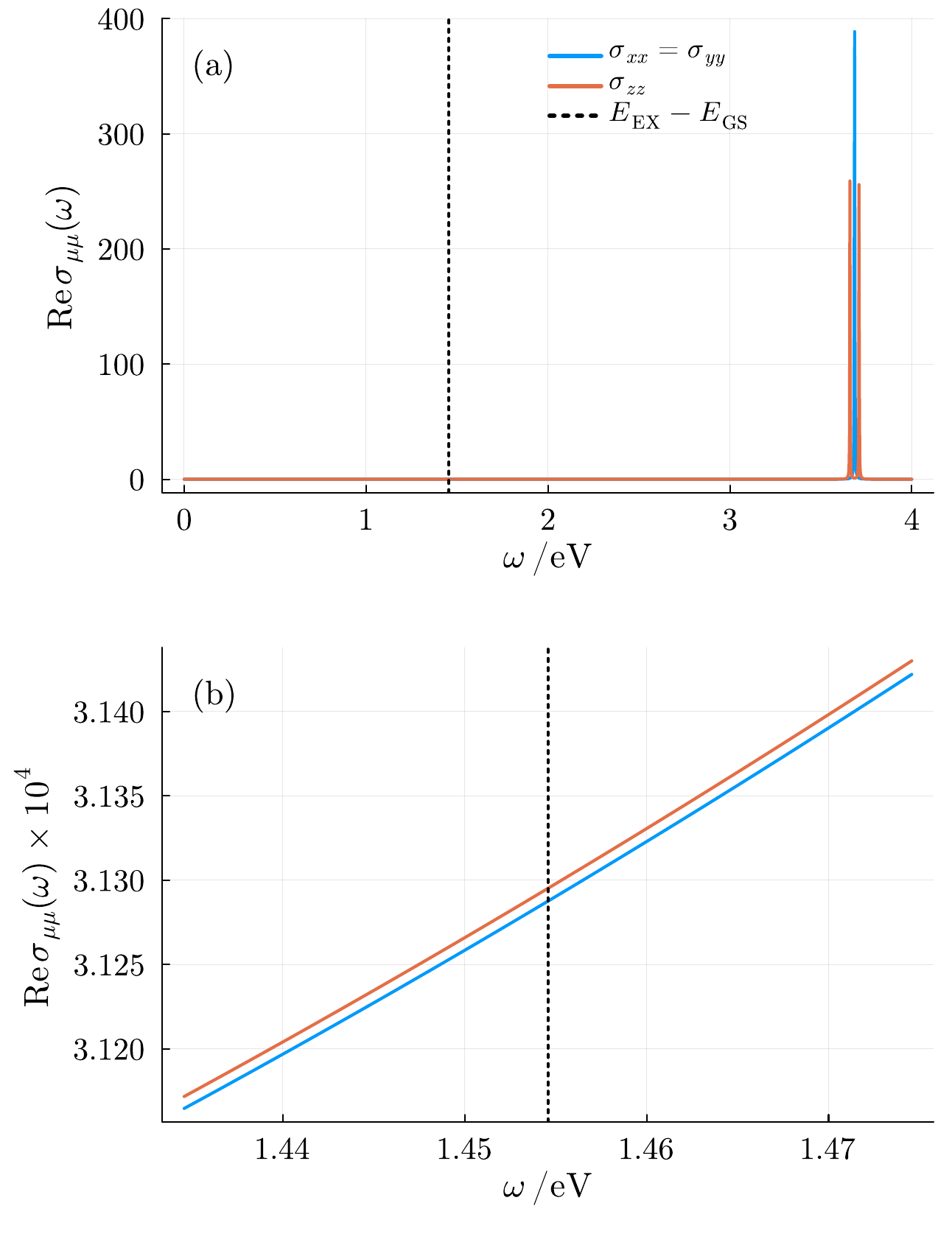}
	\caption{
		Real part of the optical conductivity $\Re\sigma_{\mu\mu}(\omega)$ calculated by regarding $\ket*{\widetilde{\mathrm{GS}},0}$ as the ground state for the single-cluster NiS$_6$ model with the Hamiltonian \eqref{eq:single_cluster_oh_hamiltonian}.
        Here  $h_{\mathrm{eff}}=0.05\,\mathrm{eV}$ and $\eta=10^{-3}\,\mathrm{eV}$.
		Panel (a) shows the spectrum over $0\leq\omega\leq4\,\mathrm{eV}$, and panel (b) shows an enlarged view around the exciton energy.
		The vertical line indicates the exciton excitation energy. 
	}
	\label{fig:single_cluster_oh_optical_conductivity}
\end{figure}

Figure~\ref{fig:single_cluster_oh_optical_conductivity} presents the resulting optical conductivity spectrum for $h_{\mathrm{eff}}=0.05\,\mathrm{eV}$.
Panel (a) provides an overview over the range $0\leq\omega\leq4\,\mathrm{eV}$, where the spectrum is dominated by the prominent charge-transfer peak near $3.7\,\mathrm{eV}$ corresponding to the transfer of a hole from $e_{g}$ orbitals to ligand $p$ orbitals.
The magnetic-field pattern leaves the $x$ and $y$ directions equivalent, leading to $\sigma_{xx}(\omega)=\sigma_{yy}(\omega)$. By contrast, it distinguishes the $z$ direction, so $\sigma_{zz}(\omega)$ is generally different from the other two components.
To examine the response near the exciton energy, panel (b) enlarges the corresponding frequency window, with the vertical line marking $E_{\widetilde{\mathrm{EX}}}-E_{\widetilde{\mathrm{GS}},0}\simeq1.454\,\mathrm{eV}$.
Although the exciton matrix element is finite, no discernible peak appears at this energy even in the enlarged spectrum.
This is because the squared matrix element $\abs*{\mel*{\widetilde{\mathrm{EX}}}{J_\mu}{\widetilde{\mathrm{GS}},0}}^2$ scales as $h_{\mathrm{eff}}^6$ and is strongly suppressed at small $h_{\mathrm{eff}}$.
For the parameters used here, the exciton contribution therefore remains too small to be resolved against the background.

\section{Enhancement of the optical response by symmetry lowering}
\label{sec:single_cluster_d3d_activation}
The scenario discussed in Sec.~\ref{sec:single_cluster_oh_activation} makes the exciton optically active, but the resulting response remains weak because the leading optical matrix element scales as $h_{\mathrm{eff}}^3$.
In this section, we show that describing the trigonal distortion by a reduction of the local symmetry from $O_h$ to $D_{3d}$ allows a contribution linear in $h_{\mathrm{eff}}$ and thereby enhances the optical response.

\subsection{Cluster model with trigonal distortion}
\label{sec:single_cluster_d3d_model}

\begin{figure}[tb]
	\centering
	\includegraphics[width=0.8\linewidth]{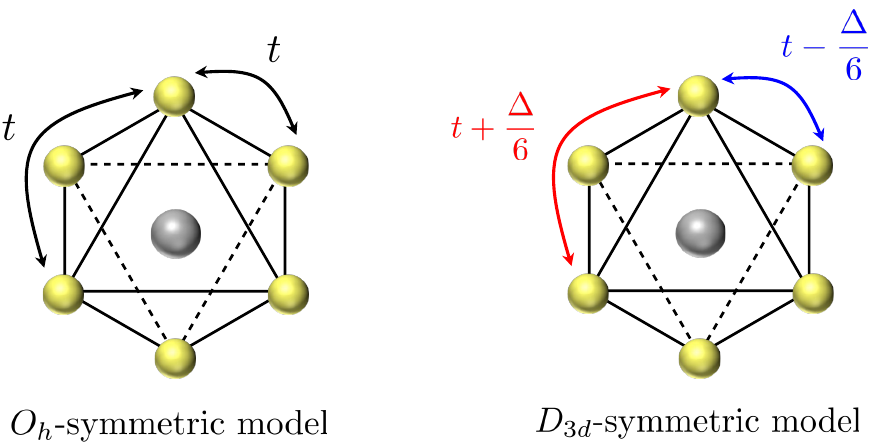}
	\caption{Schematic picture of the NiS$_6$ cluster with the octahedral symmetry $O_{h}$ (left panel) and the trigonal symmetry $D_{3d}$ (right panel).
	Under the $O_{h}$ symmetry, two hopping amplitudes shown in the left panel are equivalent, while under the $D_{3d}$ symmetry, they become inequivalent.}
	\label{fig:single_cluster_oh_to_d3d_hopping_anisotropy}
\end{figure}

Experimental studies have shown that the NiS$_6$ octahedron is trigonally distorted and that its local multiplet structure is well described by $D_{3d}$ symmetry \cite{kim2019suppression,discala2024elucidating}.
Motivated by these observations, we lower the symmetry of the idealized cluster model from $O_h$ to $D_{3d}$.
Within our $e_g$--ligand model, we represent this symmetry lowering by a $D_{3d}$-symmetric modification of the hopping between the ligand $p$ orbitals.
The resulting inequivalence of the ligand--ligand hopping amplitudes is illustrated in Fig.~\ref{fig:single_cluster_oh_to_d3d_hopping_anisotropy}.
We describe this hopping anisotropy by the term
\begin{align}
	&H_{\mathrm{trig}} = \frac{\Delta}{3} \sum_{\sigma} \Big(p_{T_{1u}^{(1)} \sigma}^{\dag} p_{T_{1u}^{(2)} \sigma} + p_{T_{1u}^{(2)} \sigma}^{\dag} p_{T_{1u}^{(3)} \sigma} \notag \\
		&\hspace{3.5cm} + p_{T_{1u}^{(3)} \sigma}^{\dag} p_{T_{1u}^{(1)} \sigma} + \mathrm{H.c.} \Big).
\end{align}
Here, $\Delta$ parametrizes the strength of the effective trigonal splitting.
Including both this term and the magnetic mean field, the single-cluster Hamiltonian becomes
\begin{align}
	H' = H_{0} + H_{\mathrm{trig}} + H_{\mathrm{mag}}.
	\label{eq:single_cluster_d3d_hamiltonian}
\end{align}
In the $O_h$ limit, the $T_{1u}^{(1)}$, $T_{1u}^{(2)}$, and $T_{1u}^{(3)}$ orbitals form a threefold-degenerate manifold transforming as the $T_{1u}$ irreducible representation.
Under $D_{3d}$, this $O_h$ representation decomposes as $T_{1u} \to A_{2u} \oplus E_u$, and $H_{\mathrm{trig}}$ accordingly splits the manifold into a nondegenerate $A_{2u}$ level and a doubly degenerate $E_u$ level.

\subsection{Group-theoretical analysis}
\label{sec:single_cluster_d3d_group_theory}

We now repeat the perturbative group-theoretical analysis of Sec.~\ref{sec:single_cluster_oh_group_theory} to determine how the trigonal distortion modifies the optical selection rules.
The perturbative expansion in $h_{\mathrm{eff}}$ remains unchanged, and an $n$th-order contribution $M_m^{(n)}$ can be nonzero only if the product in Eq.~\eqref{eq:perturbative_symmetry_product} contains the identity representation.
The difference from the $O_h$ model lies in the unperturbed Hamiltonian, which is now $H_0+H_{\mathrm{trig}}$ and has $D_{3d}$ point-group symmetry.

The transformation properties of the relevant states and operators under $D_{3d}$ and spin rotations are summarized in Table~\ref{tab:single_cluster_d3d_representations}.
Because $H_{\mathrm{trig}}$ is spin independent, the $\mathrm{SU}(2)$ transformation properties remain the same as in the $O_h$ model.
The spatial transformation properties, by contrast, must be expressed in terms of the irreducible representations of $D_{3d}$.
In particular, the $T_{1u}$ representation of $O_h$ decomposes under $D_{3d}$ as $T_{1u}\to A_{2u}\oplus E_u$.
The current operator therefore transforms as $A_{2u} \oplus E_{u}$.
For $H_{\mathrm{mag}}$, the $A_{1g}$ and $E_g$ components retain their symmetry labels under $D_{3d}$.
Of the two representations arising from its $T_{1u}$ component, however, only $E_u$ is present for the magnetic-field pattern in Eq.~\eqref{eq:magnetic_field_pattern}, while the $A_{2u}$ component vanishes.
Thus, $H_{\mathrm{mag}}$ contains only $A_{1g}$, $E_g$, and $E_u$ components.

\begin{table}[htbp]
	\centering
	\caption{
		Transformation properties of the states and operators in Eq.~\eqref{eq:perturbative_symmetry_product} under $D_{3d}$ and spin rotations.
	}
	\begin{ruledtabular}
	\begin{tabular}{ccc}
		& $D_{3d}$ symmetry & $\mathrm{SU}(2)$ symmetry \\
		\colrule
		$\ket{\mathrm{GS},m}$
		& $A_{2g}$ & $S=1$, $S^z=m$ \\
		$\ket{\mathrm{EX}}$
		& $A_{1g}$ & $S=0$, $S^z=0$ \\
		$J_\mu$
		& $A_{2u}\oplus E_u$ & rank $0$, $q=0$ \\
		$H_{\mathrm{mag}}$
		& $A_{1g}\oplus E_g\oplus E_u$
		& rank $1$, $q=0$
	\end{tabular}
	\end{ruledtabular}
	\label{tab:single_cluster_d3d_representations}
\end{table}

Substituting the representations in Table~\ref{tab:single_cluster_d3d_representations} into Eq.~\eqref{eq:perturbative_symmetry_product} gives the selection rules for $M_0^{(n)}$ summarized in Table~\ref{tab:single_cluster_d3d_selection_rules}.
Since the spin transformation properties are unchanged, the spin selection rules are the same as in the $O_h$ model.
They allow the first- and third-order contributions but forbid the second-order contribution.

The point-group constraint changes when the symmetry is lowered from $O_h$ to $D_{3d}$.
In particular, the reduced symmetry removes the constraint that forbids $M_0^{(1)}$ in the $O_h$ model.
The $D_{3d}$ point-group symmetry does not forbid any of the first three orders shown in Table~\ref{tab:single_cluster_d3d_selection_rules}.
Combining the point-group and spin constraints shows that $M_0^{(1)}$ is the leading contribution allowed by symmetry.
The optical matrix element is therefore expected to depend linearly on $h_{\mathrm{eff}}$ in the small-field regime.
A detailed derivation of these selection rules is given in Appendix~\ref{sec:appendix_group_theory}.

\begin{table}[htbp]
	\centering
	\caption{
		Point-group and spin selection rules for $M_0^{(n)}$ in the $D_{3d}$ model.
		Here, ``allowed'' means that the contribution is not forbidden by the corresponding symmetry.
	}
	\begin{ruledtabular}
	\begin{tabular}{cccc}
		& $D_{3d}$ symmetry & $\mathrm{SU}(2)$ symmetry
		& Combined result \\
		\colrule
		$M_0^{(1)}$ & allowed & allowed
		& $M_0^{(1)}$ can be nonzero \\
		$M_0^{(2)}$ & allowed & forbidden
		& $M_0^{(2)}=0$ \\
		$M_0^{(3)}$ & allowed & allowed
		& $M_0^{(3)}$ can be nonzero
	\end{tabular}
	\end{ruledtabular}
	\label{tab:single_cluster_d3d_selection_rules}
\end{table}

\subsection{Numerical results}
\label{sec:single_cluster_d3d_numerical_results}

\begin{figure}[tb]
	\centering
	\includegraphics[width=0.9\linewidth]{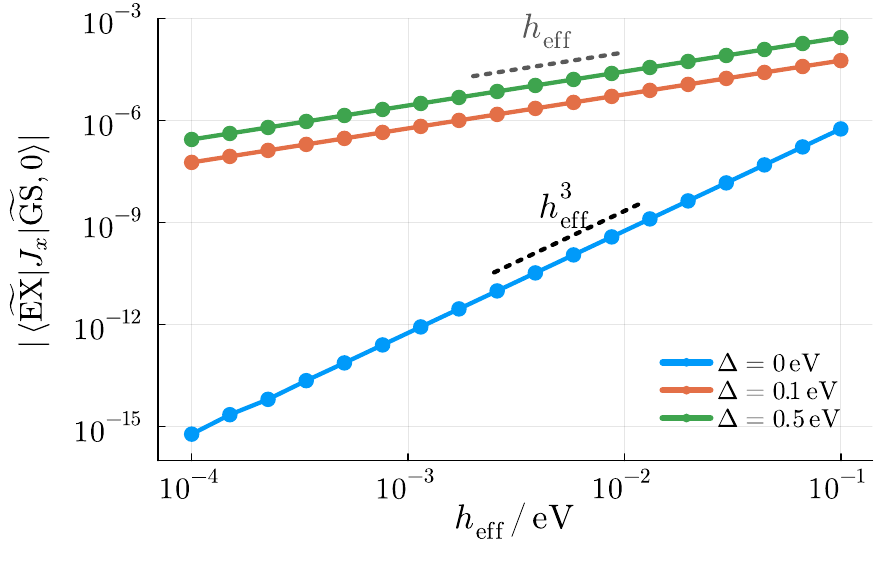}
	\caption{
		Absolute value of the matrix element $\mel*{\widetilde{\mathrm{EX}}}{J_x}{\widetilde{\mathrm{GS}},0}$ for the single-cluster NiS$_6$ model with trigonal distortion in Eq.~\eqref{eq:single_cluster_d3d_hamiltonian} as a function of the effective magnetic field $h_{\mathrm{eff}}$.
		For $\Delta=0$, the matrix element follows the cubic scaling expected in the $O_h$ limit.
		For nonzero $\Delta$, the lowering of the cluster symmetry to $D_{3d}$ allows a contribution linear in $h_{\mathrm{eff}}$.
	}
	\label{fig:single_cluster_d3d_matrix_element}
\end{figure}

We now verify the group-theoretical prediction by evaluating the optical matrix element $\mel*{\widetilde{\mathrm{EX}}}{J_x}{\widetilde{\mathrm{GS}},0}$ through exact diagonalization of the Hamiltonian \eqref{eq:single_cluster_d3d_hamiltonian}.
Figure~\ref{fig:single_cluster_d3d_matrix_element} shows the absolute value of this matrix element as a function of the effective magnetic field $h_{\mathrm{eff}}$.
For $\Delta=0$, the unperturbed cluster has $O_h$ symmetry, and the matrix element scales as $h_{\mathrm{eff}}^3$ in the small-field regime, in agreement with the group-theoretical analysis.
For nonzero $\Delta$, the symmetry of the unperturbed cluster is lowered to $D_{3d}$, and the leading dependence becomes linear in $h_{\mathrm{eff}}$, reflecting the first-order contribution allowed by the reduced point-group symmetry.
At $h_{\mathrm{eff}}=0.05\,\mathrm{eV}$, the trigonal distortion enhances the matrix element by two to three orders of magnitude relative to the $O_h$ limit.

\begin{figure}[tbp]
	\centering
	\includegraphics[width=0.9\linewidth]{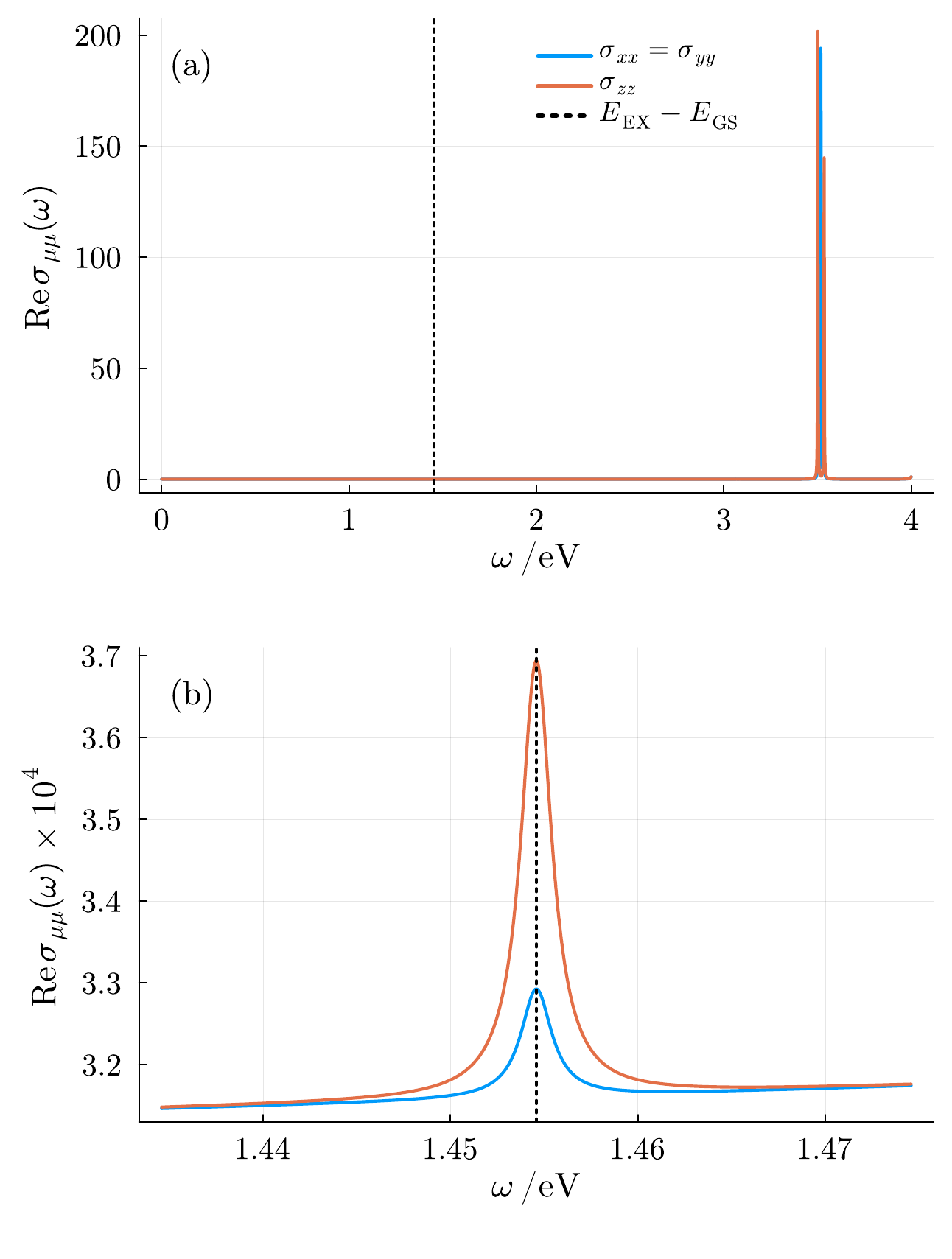}
	\caption{
		Real part of the optical conductivity $\Re\sigma_{\mu\mu}(\omega)$ for the single-cluster NiS$_6$ model with trigonal distortion described by the Hamiltonian \eqref{eq:single_cluster_d3d_hamiltonian}.
		Here, $\Delta=0.5\,\mathrm{eV}$.
		Panel (a) shows the spectrum over $0\leq\omega\leq4\,\mathrm{eV}$, and panel (b) shows an enlarged view around the exciton energy.
		The effective magnetic field and broadening factor are $h_{\mathrm{eff}}=0.05\,\mathrm{eV}$ and $\eta=10^{-3}\,\mathrm{eV}$, respectively.
	}
	\label{fig:single_cluster_d3d_optical_conductivity}
\end{figure}

To examine how the enhanced matrix element appears in the optical spectrum, we calculate the real part of the optical conductivity $\Re\sigma_{\mu\mu}(\omega)$ for the Hamiltonian \eqref{eq:single_cluster_d3d_hamiltonian}.
The resulting spectra are shown in Fig.~\ref{fig:single_cluster_d3d_optical_conductivity}.
Panel (a) displays the spectrum over $0\leq\omega\leq4\,\mathrm{eV}$, while panel (b) shows an enlarged view around the exciton energy, as in Fig.~\ref{fig:single_cluster_oh_optical_conductivity}.
As in the $O_h$ calculation, we regard the low-lying state $\ket*{\widetilde{\mathrm{GS}},0}$ as the ground state and use the broadening factor $\eta=10^{-3}\,\mathrm{eV}$.
In the present model, the $x$ and $y$ directions still remain equivalent, resulting in $\sigma_{xx}(\omega)=\sigma_{yy}(\omega)$.
The enhanced transition matrix element makes the exciton contribution distinguishable from the smooth spectral background in Fig.~\ref{fig:single_cluster_d3d_optical_conductivity}(b), in contrast to the $O_h$ result.

\section{Optical response in the two-cluster model}
\label{sec:two_cluster_model}
Thus far, our single-cluster analysis has focused on optical transitions from the low-energy triplet state with $S^z=0$ to the exciton.
However, as discussed in Sec.~\ref{sec:single_cluster_oh_activation}, the true ground state of an isolated NiS$_6$ cluster has $S^z=1$, and $S^z$ conservation forces its optical matrix element to the exciton to vanish.
In this section, we resolve this issue using a two-cluster model and show that the inter-cluster exchange interaction produces a finite optical matrix element between the true ground state and the exciton states.

\subsection{Two-cluster model}
\label{sec:two_cluster_hamiltonian}

\begin{figure}[tb]
	\centering
	\includegraphics[width=0.7\linewidth]{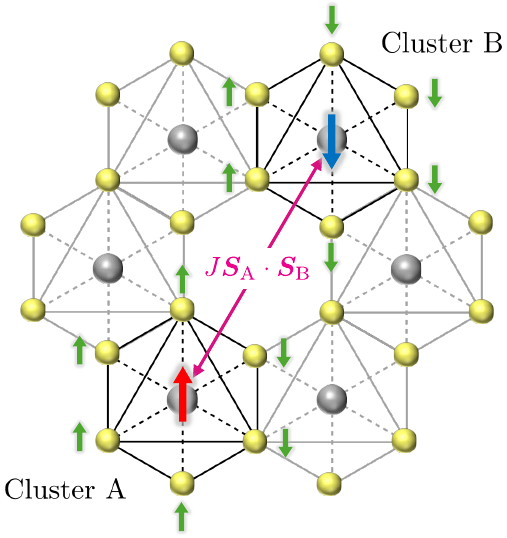}
	\caption{
		Schematic illustration of the two-cluster model.
		The red and blue arrows denote the Ni moments in clusters A and B, respectively, while the green arrows indicate the orbital-dependent effective magnetic fields acting on the ligand $p$ orbitals.
		The two Ni spins are coupled by the antiferromagnetic exchange interaction $J$.
	}
	\label{fig:two_cluster_model}
\end{figure}

We consider a model consisting of two NiS$_6$ clusters, denoted by A and B, as shown in Fig.~\ref{fig:two_cluster_model}.
The central Ni sites of the two clusters are located at opposite vertices of a hexagon in the honeycomb lattice.
This choice is motivated by the previous work showing that the antiferromagnetic correlations across a honeycomb hexagon are particularly strong \cite{autieri2022limited}.
The two-cluster Hamiltonian is
\begin{align}
	H_{\text{two-cluster}}
	={}&
	H_{\mathrm{A}} \otimes I_{\mathrm{B}}
	+I_{\mathrm{A}} \otimes H_{\mathrm{B}}
	+J\sum_{\mu=x,y,z}
	S_{\mathrm{A}}^\mu \otimes S_{\mathrm{B}}^\mu,
	\label{eq:two_cluster_hamiltonian}
\end{align}
where $H_\alpha$ and $I_\alpha$ are the single-cluster Hamiltonian and identity operator for cluster $\alpha=\mathrm{A}, \mathrm{B}$, respectively.
The operator $S_\alpha^\mu$ denotes the $\mu$ component of the Ni spin in cluster $\alpha$, and $J>0$ is the antiferromagnetic exchange coupling between the two Ni spins.

Each single-cluster Hamiltonian contains the reference term $H_0$, the trigonal term $H_{\mathrm{trig}}$, and a local magnetic mean-field term,
\begin{align}
	H_{\mathrm{A}} &= H_0+H_{\mathrm{trig}}+H_{\mathrm{mag}}^{(\mathrm{A})},\\
	H_{\mathrm{B}} &= H_0+H_{\mathrm{trig}}+H_{\mathrm{mag}}^{(\mathrm{B})}.
\end{align}
The zigzag magnetic background produces different orbital-dependent effective fields in clusters A and B, as indicated by the green arrows in Fig.~\ref{fig:two_cluster_model}.
The corresponding magnetic terms are
\begin{align}
	H_{\mathrm{mag}}^{(\mathrm{A})}
	&=
	h_{\mathrm{eff}}\big(
	S_{p_{x-}}^z+S_{p_{y+}}^z-S_{p_{x+}}^z
	\notag\\
	&\hspace{2.1cm}
	-S_{p_{y-}}^z-S_{p_{z+}}^z-S_{p_{z-}}^z
	\big),\\
	H_{\mathrm{mag}}^{(\mathrm{B})}
	&=
	h_{\mathrm{eff}}\big(
	S_{p_{x-}}^z+S_{p_{y+}}^z-S_{p_{x+}}^z
	\notag\\
	&\hspace{2.1cm}
	-S_{p_{y-}}^z+S_{p_{z+}}^z+S_{p_{z-}}^z
	\big).
\end{align}

When $J=0$, the model reduces to two independent single-cluster systems.
The opposite magnetic environments of clusters A and B select $\ket*{\widetilde{\mathrm{GS}},1}$ and $\ket*{\widetilde{\mathrm{GS}},-1}$ as their respective ground states.
The ground state of the decoupled two-cluster system is consequently the product state
\begin{align}
	\ket*{\widetilde{\mathrm{GS}},1}
	\otimes
	\ket*{\widetilde{\mathrm{GS}},-1}.
\end{align}
For each isolated cluster, conservation of $S^z$ forbids the optical transition from its true ground state with $S^z=\pm1$ to the exciton with $S^z=0$, as discussed in Sec.~\ref{sec:single_cluster_oh_numerical_results}.
An optical excitation acting on either cluster therefore cannot connect the product ground state to the one-exciton product states
\begin{align}
	\ket*{\widetilde{\mathrm{EX}}}
	\otimes
	\ket*{\widetilde{\mathrm{GS}},-1},
	\qquad
	\ket*{\widetilde{\mathrm{GS}},1}
	\otimes
	\ket*{\widetilde{\mathrm{EX}}}.
\end{align}

The situation changes when a finite inter-cluster exchange coupling is introduced.
The exchange interaction generates quantum spin fluctuations between the two clusters, and the ground state acquires components proportional to $\ket*{\widetilde{\mathrm{GS}},1}\otimes\ket*{\widetilde{\mathrm{GS}},-1}$, $\ket*{\widetilde{\mathrm{GS}},0}\otimes\ket*{\widetilde{\mathrm{GS}},0}$, and $\ket*{\widetilde{\mathrm{GS}},-1}\otimes\ket*{\widetilde{\mathrm{GS}},1}$.
In particular, the $\ket*{\widetilde{\mathrm{GS}},0}\otimes\ket*{\widetilde{\mathrm{GS}},0}$ component can be optically excited to $\ket*{\widetilde{\mathrm{EX}}}\otimes\ket*{\widetilde{\mathrm{GS}},0}$ or $\ket*{\widetilde{\mathrm{GS}},0}\otimes\ket*{\widetilde{\mathrm{EX}}}$.
The inter-cluster exchange interaction can therefore produce a finite optical matrix element between the true ground state of the coupled system and the exciton states.

\subsection{Numerical results}
\label{sec:two_cluster_numerical_results}
To test the above argument, we numerically diagonalize the Hamiltonian \eqref{eq:two_cluster_hamiltonian} and examine the optical response associated with the exciton.
We set the inter-cluster exchange coupling to $J=15\,\mathrm{meV}$, which is close to the value estimated in Ref.~\cite{autieri2022limited}.
For this value of $J$, the two-cluster model has a unique ground state $\ket*{\Psi}$ with total $S^z=0$.

Extending the single-cluster optical-matrix-element analysis of Secs.~\ref{sec:single_cluster_oh_numerical_results} and \ref{sec:single_cluster_d3d_numerical_results} to the two-cluster model requires two additional steps, namely identifying the exciton states among the eigenstates of the coupled system and defining the optical matrix element for transitions into the resulting exciton subspace.
The identification of the exciton states is most transparent in the decoupled limit $J=0$.
In this limit, the product states
\begin{align}
    \ket*{\mathrm{EX}^{(0)}_{\mathrm{A}},m}
    &\coloneqq
    \ket*{\widetilde{\mathrm{EX}}}
    \otimes
    \ket*{\widetilde{\mathrm{GS}},m},\\
    \ket*{\mathrm{EX}^{(0)}_{\mathrm{B}},m}
    &\coloneqq
    \ket*{\widetilde{\mathrm{GS}},m}
    \otimes
    \ket*{\widetilde{\mathrm{EX}}}
    \label{eq:two_cluster_one_exciton_states}
\end{align}
are eigenstates of the two-cluster Hamiltonian for each $m=1,0,-1$.
They describe an exciton localized in cluster A or B, respectively, and together span the one-exciton subspace with total $S^z=m$.
For finite $J$, however, these states are generally no longer eigenstates of the Hamiltonian, and the exciton-like eigenstates obtained by numerical diagonalization may contain components along both $\ket*{\mathrm{EX}^{(0)}_{\mathrm{A}},m}$ and $\ket*{\mathrm{EX}^{(0)}_{\mathrm{B}},m}$.
To identify the exciton states, we therefore define the projection weight of an exact eigenstate $\ket*{n,m}$ as
\begin{align}
    W_n^{(m)}
    =
    \abs*{\braket*{\mathrm{EX}^{(0)}_{\mathrm{A}},m}{n,m}}^2
    +
    \abs*{\braket*{\mathrm{EX}^{(0)}_{\mathrm{B}},m}{n,m}}^2.
    \label{eq:two_cluster_exciton_projection_weight}
\end{align}
The weight $W_n^{(m)}$ ranges from zero to unity and approaches unity when $\ket*{n,m}$ lies predominantly within the one-exciton subspace.
We therefore identify eigenstates with $W_n^{(m)}$ close to unity as the exciton states of the coupled system.

Because the ground state $\ket*{\Psi}$ has total $S^z=0$ and the two-cluster current operator $J_x$ conserves total $S^z$, optical transitions from $\ket*{\Psi}$ can reach only exciton states in the $m=0$ sector.
Thus, we identify the corresponding relevant exciton states as the two exact eigenstates whose projection weights $W_n^{(0)}$ are close to unity and denote them by $\ket*{\mathrm{EX}_1}$ and $\ket*{\mathrm{EX}_2}$.

Because the optical weight is shared by these two exciton states, we define the combined optical matrix element
\begin{align}
    M_{\mathrm{2cl}}
    =
    \qty(
        \sum_{i=1}^{2}
        \abs*{\mel*{\mathrm{EX}_i}{J_x}{\Psi}}^2
    )^{1/2}.
\end{align}
Because $M_{\mathrm{2cl}}^2$ gives the total transition strength from the true ground state to the two-state exciton subspace, $M_{\mathrm{2cl}}$ provides the natural two-cluster counterpart of the single-cluster matrix element $\mel*{\widetilde{\mathrm{EX}}}{J_x}{\widetilde{\mathrm{GS}},0}$.

\begin{figure}[tb]
    \centering
    \includegraphics[width=0.9\linewidth]{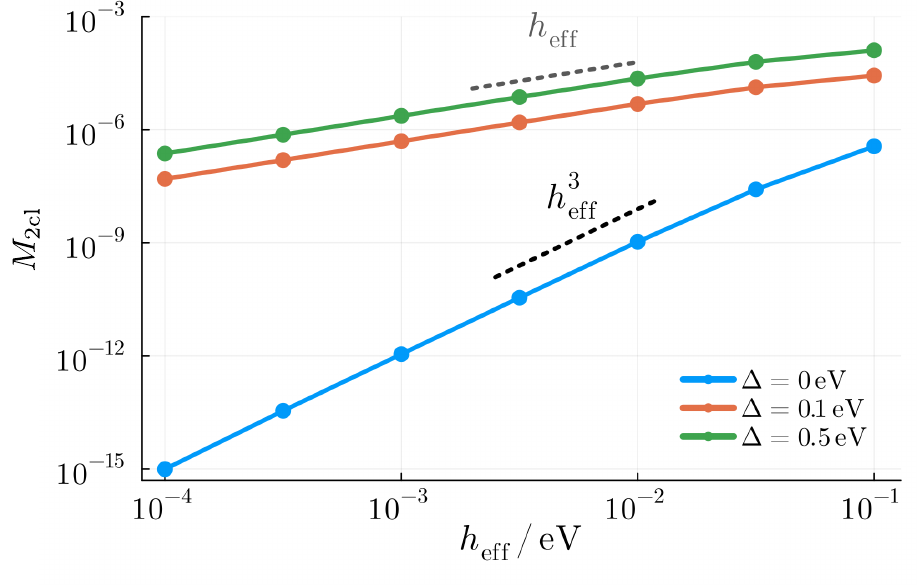}
    \caption{
        Generalized optical matrix element $M_{\mathrm{2cl}}$ as a function of the effective magnetic field $h_{\mathrm{eff}}$.
        For $\Delta=0$, $M_{\mathrm{2cl}}$ exhibits cubic scaling, whereas for nonzero $\Delta$, the $D_{3d}$ symmetry lowering allows a contribution linear in $h_{\mathrm{eff}}$.
    }
    \label{fig:two_cluster_matrix_element}
\end{figure}

Figure~\ref{fig:two_cluster_matrix_element} shows the dependence of $M_{\mathrm{2cl}}$ on the effective magnetic field $h_{\mathrm{eff}}$.
Consistent with the single-cluster results, $M_{\mathrm{2cl}}$ becomes finite for nonzero $h_{\mathrm{eff}}$ and scales as $h_{\mathrm{eff}}^3$ in the small-field regime when $\Delta=0$.
For nonzero $\Delta$, by contrast, its leading dependence becomes linear in $h_{\mathrm{eff}}$, demonstrating that the symmetry lowering also enhances the optical response in the two-cluster model.

\begin{figure}[tb]
    \centering
    \includegraphics[width=0.9\linewidth]{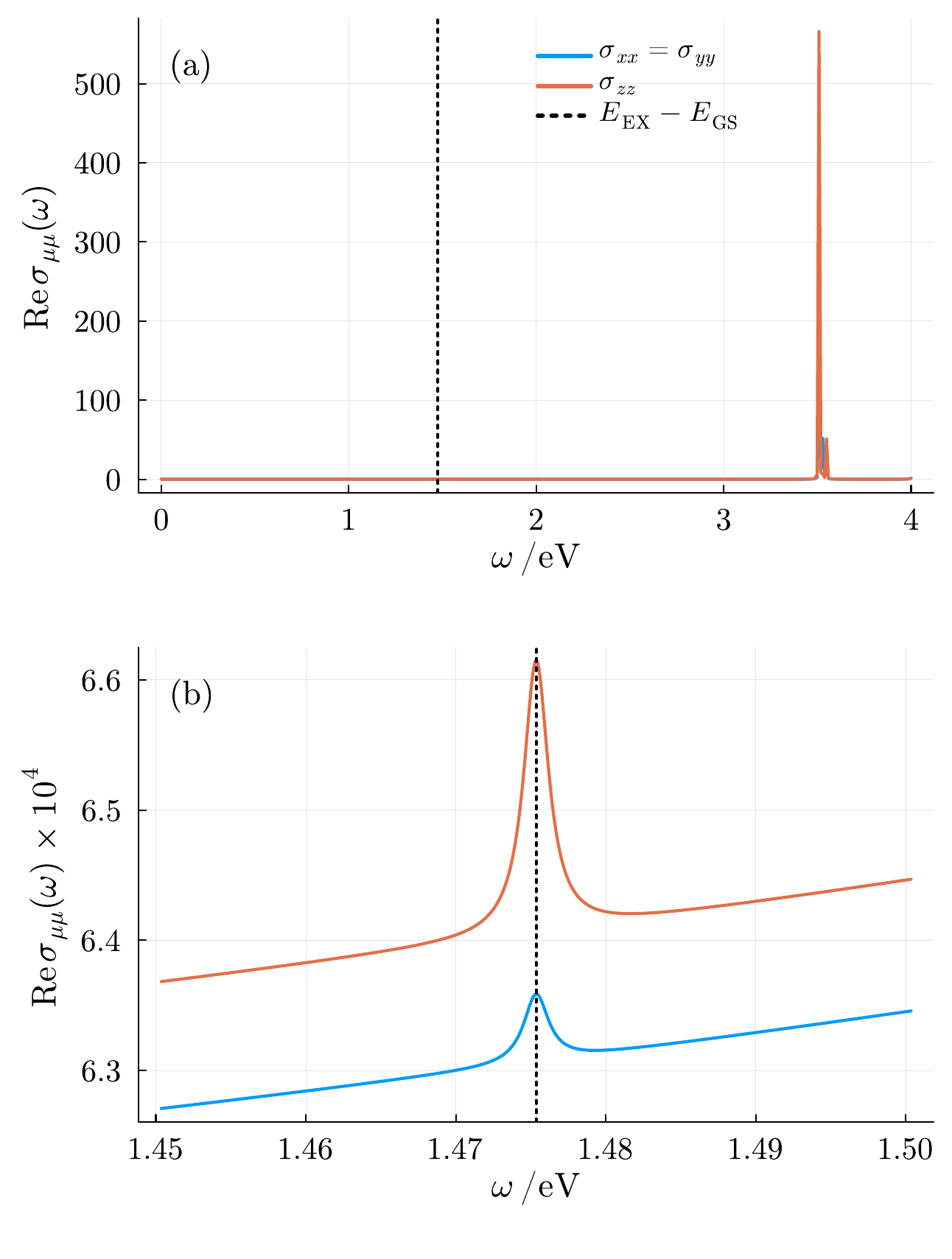}
    \caption{
        Longitudinal optical-conductivity components $\Re\sigma_{\mu\mu}(\omega)$, with $\mu=x,y,z$, for the two-cluster model described by Eq.~\eqref{eq:two_cluster_hamiltonian}.
        Panel (a) shows the spectrum over $0\leq\omega\leq4\,\mathrm{eV}$, and panel (b) shows an enlarged view around the exciton energy.
        The two exciton states are degenerate within numerical accuracy, and their common excitation energy of approximately $1.475\,\mathrm{eV}$ is indicated by the dotted vertical line.
        The parameters are $J=15\,\mathrm{meV}$, $\Delta=0.5\,\mathrm{eV}$, $h_{\mathrm{eff}}=0.05\,\mathrm{eV}$, and $\eta=10^{-3}\,\mathrm{eV}$.
    }
    \label{fig:two_cluster_d3d_optical_conductivity}
\end{figure}

We next calculate the longitudinal optical-conductivity components $\Re\sigma_{\mu\mu}(\omega)$, with $\mu=x,y,z$, for the two-cluster model, as shown in Fig.~\ref{fig:two_cluster_d3d_optical_conductivity}.
For this calculation, we set $h_{\mathrm{eff}}=0.05\,\mathrm{eV}$, $\Delta=0.5\,\mathrm{eV}$, $J=15\,\mathrm{meV}$, and $\eta=10^{-3}\,\mathrm{eV}$.
With these parameters, the two exciton states $\ket*{\mathrm{EX}_1}$ and $\ket*{\mathrm{EX}_2}$ are degenerate within numerical accuracy, with a common excitation energy of approximately $1.475\,\mathrm{eV}$.
The relation $\sigma_{xx}(\omega)=\sigma_{yy}(\omega)$, with $\sigma_{zz}(\omega)$ generally different, is preserved in the two-cluster model.

In the single-cluster calculations, we used the low-lying state $\ket*{\widetilde{\mathrm{GS}},0}$ as the initial state because the optical matrix element from the true ground state to the exciton vanishes.
In the two-cluster model, by contrast, the optical conductivity can be evaluated directly from the true ground state $\ket*{\Psi}$ as
\begin{align}
    \Re \sigma_{\mu\mu}(\omega>0)
    =
    \sum_{n\neq\Psi}
    \frac{\abs*{\mel*{n}{J_\mu}{\Psi}}^2}{E_n-E_\Psi}
    \frac{\eta}
    {(\omega-E_n+E_\Psi)^2+\eta^2},
\end{align}
where the sum runs over the excited eigenstates $\ket*{n}$ with energies $E_n$, and $E_\Psi$ is the ground-state energy.
The spectrum retains the overall structure found in the single-cluster model, with a charge-transfer peak near $3.7\,\mathrm{eV}$ and a distinct excitonic feature near $1.475\,\mathrm{eV}$ that is resolved from the smooth background.
This result shows that the inter-cluster exchange generates the quantum fluctuations required to produce finite optical matrix elements between the true ground state and the exciton states.

\section{Summary and outlook}
\label{sec:summary_and_outlook}

In this work, we investigated the optical response of the many-body exciton in NiPS$_3$.
We first considered a reference single-cluster NiS$_6$ model and showed that the matrix element governing one-photon absorption and emission vanishes owing to spin and parity selection rules.
The reference model therefore cannot account for the experimentally observed peak of the exciton.

To resolve this discrepancy, we incorporated the influence of the surrounding zigzag antiferromagnetic order into the single-cluster model at the mean-field level.
The resulting symmetry breaking relaxes the spin and parity selection rules and makes the optical matrix element finite.
This mechanism also explains the reason why the exciton is optically active only below the N\'{e}el temperature, where the zigzag order is established.
Our group-theoretical analysis predicts that the leading nonzero contribution to the transition amplitude scales as $h_{\mathrm{eff}}^3$, and the same scaling is confirmed by exact diagonalization.
Although the magnetic mean field activates the transition, its cubic dependence on $h_{\mathrm{eff}}$ strongly suppresses the optical response in the ideal $O_h$ limit, leaving no discernible exciton peak in the calculated optical-conductivity spectrum.

As a possible mechanism for enhancing this weak exciton response, we next examined the effect of the trigonal distortion present in the actual crystal structure.
Lowering the local symmetry from $O_h$ to $D_{3d}$ modifies the selection rules and allows a contribution linear in $h_{\mathrm{eff}}$.
As a result, the optical matrix element is enhanced by two to three orders of magnitude relative to the $O_h$ limit, producing a visible exciton peak in the optical-conductivity spectrum.

Finally, we extended the analysis to a two-cluster model to incorporate the inter-cluster exchange interaction.
In the single-cluster model, conservation of $S^z$ prohibits the transition from the true ground state with $S^z=1$ to the singlet exciton.
The single-cluster optical response was therefore evaluated from the low-lying triplet state with $S^z=0$.
In the two-cluster model, the exchange interaction generates quantum spin fluctuations that mix configurations containing $S^z=0$ components into the ground state.
This mixing produces a finite optical matrix element between the true ground state of the coupled system and the exciton states.
Exact diagonalization confirms that the resulting optical-conductivity spectrum exhibits a clear exciton peak.

Although our scenario accounts for the optical visibility of the many-body exciton in optical absorption and photoluminescence, several experimentally observed properties remain beyond the scope of the present model.
One such issue concerns the pronounced polarization anisotropy of the exciton.
Polarization-resolved measurements have revealed that the exciton photoluminescence is nearly fully linearly polarized and that the corresponding optical response exhibits pronounced linear dichroism \cite{hwangbo2021highly,wang2021spin,kim2023anisotropic}.
Moreover, magneto-photoluminescence experiments have shown that the emission-polarization axis follows the field-induced reorientation of the N\'{e}el vector \cite{song2024manipulation,wang2024unveiling}.
Together, these observations indicate that the optical dipole associated with the exciton is closely linked to the orientation of the antiferromagnetic order.
In the present model, however, spin rotations and real-space point-group operations act independently.
Thus, changing the orientation of the N\'{e}el vector by a global spin rotation leaves the optical response unchanged, and the model therefore cannot reproduce the observed dependence of the emission polarization on the spin orientation.
Incorporating spin-orbit coupling into a more complete multiorbital description, thereby coupling the spin orientation to the orbital and crystallographic degrees of freedom, will be essential for identifying the microscopic origin of this polarization dependence.

A second open issue concerns the exceptionally narrow linewidth of the exciton peak.
Several experiments have suggested that thermal and field-induced fluctuations of the antiferromagnetic order contribute to the broadening and decoherence of the exciton \cite{lee2024optical,lee2024imaging,wang2024unveiling}.
In the present calculations, however, the isolated finite-size systems have discrete spectra and contain no intrinsic relaxation or dephasing mechanisms.
The linewidths displayed in the calculated spectra are therefore determined entirely by the phenomenological broadening parameter $\eta$ and do not provide information about the intrinsic exciton lifetime or coherence.
Understanding the origin of the ultranarrow linewidth will require a treatment of larger systems, together with the coupling of the exciton to magnons, phonons, and other radiative and nonradiative decay channels.
More sophisticated numerical methods such as DMRG could facilitate such an extension and help clarify why the exciton retains a well-defined, ultranarrow spectral line at low temperatures and how its linewidth evolves as the antiferromagnetic order becomes increasingly fluctuating or disordered.

Overall, while there are several open questions regarding the exciton and its coupling to the antiferromagnetic order in NiPS$_{3}$, we provided a microscopic mechanism for the optical activation of the exciton and demonstrated that the zigzag antiferromagnetic order plays a crucial role in enabling its optical visibility.

\acknowledgments
We thank K. M. Dani, T. Kaneko, H. Miyamoto, Y. Murakami, T. Nakamoto, S. Okada, S. Takayoshi, and N. Tomoda for fruitful discussions.
This work was supported by JST FOREST (Grant No. JPMJFR2131) and JSPS KAKENHI (Grant Nos. JP24H00191, JP25H01246 and JP25H01251).
S.M. was also supported by JSPS KAKENHI (Grant No. JP26KJ1014), the Forefront Physics and Mathematics Program to Drive Transformation (FoPM), a World-Leading Innovative Graduate Study (WINGS) Program and the JSR Fellowship, the University of Tokyo.
S.I. acknowledges support from JSPS KAKENHI (Grant Nos. JP25K17343 and JP26K21749).
The computations for this work have been done using the facilities of the Supercomputer Center, the Institute for Solid State Physics, the University of Tokyo (2026-Ba-0003).

\appendix
\onecolumngrid
\section{Eigenstates of the reference cluster model}
\label{sec:appendix_eigenstates}
In this Appendix, we show the explicit forms of the states appearing in Eqs.~\eqref{eq:ground_state} and \eqref{eq:exciton_state}.
The states are classified by the number of electrons in the Ni $d$ orbitals $n$, the number of ligand holes $k$, the irreducible representation $\Gamma$ of the octahedral group $O_h$, total spin $S$, and the $z$-component of the total spin $m$. We denote these states by $\ket*{d^n\underline{L}^k,{}^{2S+1}\Gamma,m}$.
The ground-state manifold is a spin-triplet state with $A_{2g}$ symmetry, for which the relevant states are given by
\begin{align}
	&\ket*{d^{8}, {}^{3}A_{2g}, 1} =
	d_{E_{g}^{(1)} \uparrow}^{\dag}
	d_{E_{g}^{(2)} \uparrow}^{\dag}
	\ket{0}, \\
	&\ket*{d^{9}\underline{L}^{1}, {}^{3}A_{2g}, 1} =
	\frac{1}{\sqrt{2}}
  \qty(
  d^{\dag}_{E_{g}^{(1)} \uparrow}
  p^{\dag}_{E_{g}^{(2)} \uparrow}
  +
  p^{\dag}_{E_{g}^{(1)} \uparrow}
  d^{\dag}_{E_{g}^{(2)} \uparrow}
  )
  \ket*{0}, \\
	&\ket*{d^{10}\underline{L}^{2}, {}^{3}A_{2g}, 1} =
	p^{\dag}_{E_{g}^{(1)} \uparrow}
	p^{\dag}_{E_{g}^{(2)} \uparrow}
	\ket{0}, \\
	&\ket*{d^{8}, {}^{3}A_{2g}, 0} =
	\frac{1}{\sqrt{2}}
	\qty(
	d^{\dag}_{E_{g}^{(1)} \uparrow}
	d^{\dag}_{E_{g}^{(2)} \downarrow}
	+
	d^{\dag}_{E_{g}^{(1)} \downarrow}
	d^{\dag}_{E_{g}^{(2)} \uparrow})
	\ket*{0}, \\
	&\ket*{d^{9}\underline{L}^{1}, {}^{3}A_{2g}, 0}
  =
  \frac{1}{2}
  \qty(
  d^{\dag}_{E_{g}^{(1)} \uparrow}
  p^{\dag}_{E_{g}^{(2)} \downarrow}
  +
  d^{\dag}_{E_{g}^{(1)} \downarrow}
  p^{\dag}_{E_{g}^{(2)} \uparrow}
  +
  p^{\dag}_{E_{g}^{(1)} \uparrow}
  d^{\dag}_{E_{g}^{(2)} \downarrow}
  +
  p^{\dag}_{E_{g}^{(1)} \downarrow}
  d^{\dag}_{E_{g}^{(2)} \uparrow}
  )
  \ket*{0}, \\
	&\ket*{d^{10}\underline{L}^{2}, {}^{3}A_{2g}, 0} =
	\frac{1}{\sqrt{2}}
	\qty(
	p^{\dag}_{E_{g}^{(1)} \uparrow}
	p^{\dag}_{E_{g}^{(2)} \downarrow}
	+
	p^{\dag}_{E_{g}^{(1)} \downarrow}
	p^{\dag}_{E_{g}^{(2)} \uparrow})
	\ket*{0}, \\
	&\ket*{d^{8}, {}^{3}A_{2g}, -1} = d_{E_{g}^{(1)} \downarrow}^{\dag} d_{E_{g}^{(2)} \downarrow}^{\dag} \ket{0}, \\
	&\ket*{d^{9}\underline{L}^{1}, {}^{3}A_{2g}, -1} =
	\frac{1}{\sqrt{2}}
  \qty(
  d^{\dag}_{E_{g}^{(1)} \downarrow}
  p^{\dag}_{E_{g}^{(2)} \downarrow}
  +
  p^{\dag}_{E_{g}^{(1)} \downarrow}
  d^{\dag}_{E_{g}^{(2)} \downarrow}
  )
  \ket*{0}, \\
	&\ket*{d^{10}\underline{L}^{2}, {}^{3}A_{2g}, -1} =
	p^{\dag}_{E_{g}^{(1)} \downarrow}
	p^{\dag}_{E_{g}^{(2)} \downarrow}
	\ket{0}.
\end{align}
The many-body exciton state is a spin-singlet state with $A_{1g}$ symmetry, for which the relevant states are given by
\begin{align}
	&\ket*{d^{8}, {}^{1}A_{1g}, 0} =
	\frac{1}{\sqrt{2}}
  \qty(
  d^{\dag}_{E_{g}^{(1)} \uparrow}
  d^{\dag}_{E_{g}^{(1)} \downarrow}
  +
  d^{\dag}_{E_{g}^{(2)} \uparrow}
  d^{\dag}_{E_{g}^{(2)} \downarrow}
  )
	\ket{0}, \\
	&\ket*{d^{9}\underline{L}^{1}, {}^{1}A_{1g}, 0} =
	\frac{1}{2}
  \qty(
  d^{\dag}_{E_{g}^{(1)} \uparrow}
  p^{\dag}_{E_{g}^{(1)} \downarrow}
  +
  p^{\dag}_{E_{g}^{(1)} \uparrow}
  d^{\dag}_{E_{g}^{(1)} \downarrow}
  +
  d^{\dag}_{E_{g}^{(2)} \uparrow}
  p^{\dag}_{E_{g}^{(2)} \downarrow}
  +
  p^{\dag}_{E_{g}^{(2)} \uparrow}
  d^{\dag}_{E_{g}^{(2)} \downarrow}
  )
  \ket*{0}, \\
	&\ket*{d^{10}\underline{L}^{2}, {}^{1}A_{1g}, 0} =
	\frac{1}{\sqrt{2}}
	\qty(
	p^{\dag}_{E_{g}^{(1)} \uparrow}
	p^{\dag}_{E_{g}^{(1)} \downarrow}
	+
	p^{\dag}_{E_{g}^{(2)} \uparrow}
	p^{\dag}_{E_{g}^{(2)} \downarrow}
	)
    \ket*{0}.
\end{align}

\section{Derivation of optical selection rules}
\label{sec:appendix_group_theory}

In this Appendix, we show the details of the derivation of the selection rules for the optical matrix elements in Secs.~\ref{sec:single_cluster_oh_group_theory} and \ref{sec:single_cluster_d3d_group_theory}.
The common criterion is whether the product of the states and operators in Eq.~\eqref{eq:perturbative_symmetry_product} contains a nonzero component that transforms as the trivial representation.
We first apply this criterion to the spatial symmetries $O_h$ and $D_{3d}$ and then derive the spin selection rules using the spin-rotational $\mathrm{SU}(2)$ symmetry of the unperturbed Hamiltonian.

\subsection{Selection rules under $O_{h}$ symmetry}
For the $O_h$ model, the representations entering Eq.~\eqref{eq:perturbative_symmetry_product} are listed in Table~\ref{tab:single_cluster_oh_representations}.
Substituting them into the product gives the following decompositions for $n=1,2,3$:
\begin{align}
	(A_{1g} \oplus E_{g} \oplus T_{1u}) \otimes A_{1g} \otimes T_{1u} \otimes A_{2g}
	&= A_{2g} \oplus E_{g} \oplus T_{1g} \oplus T_{2g} \oplus T_{1u} \oplus 2T_{2u}, \\
		(A_{1g} \oplus E_{g} \oplus T_{1u})^{\otimes 2} \otimes A_{1g} \otimes T_{1u} \otimes A_{2g}
	&= 2A_{1g} \oplus 4A_{2g} \oplus 6E_{g} \oplus 6T_{1g} \oplus 6T_{2g} \notag \\
	&\hspace{3cm} \oplus A_{1u} \oplus A_{2u} \oplus 2E_{u} \oplus 7T_{1u} \oplus 9T_{2u}, \\
		(A_{1g} \oplus E_{g} \oplus T_{1u})^{\otimes 3} \otimes A_{1g} \otimes T_{1u} \otimes A_{2g}
		&= 15A_{1g} \oplus 19A_{2g} \oplus 34E_{g} \oplus 37T_{1g} \oplus 37T_{2g} \notag \\
	&\hspace{3cm} \oplus 9A_{1u} \oplus 9A_{2u} \oplus 18E_{u} \oplus 43T_{1u} \oplus 47T_{2u}.
\end{align}
A contribution is not forbidden by $O_h$ symmetry only when its decomposition contains the trivial representation $A_{1g}$.
Because $A_{1g}$ is absent at first order, $M_0^{(1)}$ must vanish in the $O_h$ model.
At second and third orders, $A_{1g}$ is present, so the spatial selection rule does not prohibit $M_0^{(2)}$ or $M_0^{(3)}$.

\subsection{Selection rules under $D_{3d}$ symmetry}
For the $D_{3d}$ model, the corresponding representations are given in Table~\ref{tab:single_cluster_d3d_representations}.
The same product then decomposes as
\begin{align}
		(A_{1g} \oplus E_{g} \oplus E_{u}) \otimes A_{1g} \otimes (A_{2u} \oplus E_{u}) \otimes A_{2g}
	&= A_{1g} \oplus A_{2g} \oplus 2E_{g} \oplus 2A_{1u} \oplus A_{2u} \oplus 3E_{u}, \\
		(A_{1g} \oplus E_{g}  \oplus E_{u})^{\otimes 2} \otimes A_{1g} \otimes (A_{2u} \oplus E_{u}) \otimes A_{2g}
	&= 6A_{1g} \oplus 6A_{2g} \oplus 12E_{g} \oplus 7A_{1u} \oplus 6A_{2u} \oplus 13E_{u}, \\
		(A_{1g} \oplus E_{g}  \oplus E_{u})^{\otimes 3} \otimes A_{1g} \otimes (A_{2u} \oplus E_{u}) \otimes A_{2g}
	&= 31A_{1g} \oplus 31A_{2g} \oplus 62E_{g} \oplus 32A_{1u} \oplus 31A_{2u} \oplus 63E_{u}.
\end{align}
In contrast to the $O_h$ result, the trivial representation $A_{1g}$ appears at every order shown above.
The $D_{3d}$ spatial selection rule therefore does not prohibit $M_0^{(1)}$, $M_0^{(2)}$, or $M_0^{(3)}$.

\subsection{Selection rules under $\mathrm{SU}(2)$ symmetry}

We next derive the spin selection rules associated with the product representation in Eq.~\eqref{eq:perturbative_symmetry_product}.
As shown in Table~\ref{tab:single_cluster_oh_representations}, the ground state $\ket*{\mathrm{GS},m}$ has $S=1$ and $S^z=m$, whereas the exciton state $\ket*{\mathrm{EX}}$ has $S=0$ and $S^z=0$.
Under spin rotations, the current operator $J_{\mu}$ is a scalar, whereas $H_{\mathrm{mag}}$ transforms as the $q=0$ component of a rank-1 spherical tensor.

Every term contributing to the $n$th-order coefficient $M_m^{(n)}$ contains a total of $n$ insertions of $H_{\mathrm{mag}}$.
Since the other factors in the perturbative expansion are spin scalars, they do not affect the spin selection rule.
The spin dependence of an $n$th-order term can therefore be represented schematically by
\begin{align}
    \mel{0,0}{\qty(H_{\mathrm{mag}})^n}{1,m}.
    \label{eq:nth_order_spin_structure}
\end{align}

Each factor of $H_{\mathrm{mag}}$ has $q=0$, so the product in Eq.~\eqref{eq:nth_order_spin_structure} cannot change $S^z$.
Because the exciton has $S^z=0$, the matrix element vanishes for $m\neq0$.
It is therefore sufficient to consider $M_0^{(n)}$.

Before examining the individual orders, we determine which ranks of spherical tensors can connect $\ket*{1,0}$ and $\ket*{0,0}$.
Let $T_0^{(K)}$ denote the $Q=0$ component of an irreducible spherical tensor of rank $K$.
The Wigner--Eckart theorem gives
\begin{align}
    \mel{0,0}{T_0^{(K)}}{1,0}
    =
    C^{00}_{10,K0}
    \langle 0 \Vert T^{(K)} \Vert 1 \rangle.
    \label{eq:wigner_eckart_spin_selection}
\end{align}
Here, $\langle 0 \Vert T^{(K)} \Vert 1 \rangle$ is the reduced matrix element, which is independent of the initial and final values of $S^z$ and of the tensor component $Q$.
The factor $C^{00}_{10,K0}$ is the Clebsch--Gordan coefficient.
This coefficient vanishes unless $K=1$, so $\mel*{0,0}{T_0^{(K)}}{1,0}$ is forbidden for all $K\neq1$.
It follows that $M_0^{(n)}$ is not forbidden by the spin selection rule only if the irreducible decomposition of $\qty(H_{\mathrm{mag}})^n$ contains a rank-1, $Q=0$ component.

At first order, $H_{\mathrm{mag}}$ itself transforms as a rank-1, $q=0$ tensor.
The required component is therefore present, and the spin selection rule allows $M_0^{(1)}$.

At second order, the spin-tensor structure of $\qty(H_{\mathrm{mag}})^2$ is that of the product of two rank-1, $q=0$ tensors.
Using $T_0^{(1)}$ to represent a rank-1, $q=0$ tensor, this product can be decomposed as
\begin{align}
    T_0^{(1)}T_0^{(1)}
    =
    \sum_{K=0}^{2}
    C^{K0}_{10,10}
    \qty[T^{(1)}\otimes T^{(1)}]_0^{(K)}.
    \label{eq:second_order_spin_decomposition}
\end{align}
Here, $\qty[T^{(1)}\otimes T^{(1)}]_0^{(K)}$ denotes the rank-$K$, $Q=0$ tensor obtained by coupling the two rank-1 tensors.
The relevant Clebsch--Gordan coefficients satisfy
\begin{align}
    C^{00}_{10,10}\neq0,\qquad
    C^{10}_{10,10}=0,\qquad
    C^{20}_{10,10}\neq0.
\end{align}
The decomposition in Eq.~\eqref{eq:second_order_spin_decomposition} therefore contains rank-0 and rank-2 components, but no rank-1 component.
Since tensors of neither rank can connect $\ket*{1,0}$ to $\ket*{0,0}$, the spin selection rule prohibits $M_0^{(2)}$.

At third order, the first two rank-1, $q=0$ tensors can be coupled to rank 0 because $C^{00}_{10,10}\neq0$.
This rank-0 tensor can then be coupled to the third rank-1 tensor to form a rank-1 tensor, since
\begin{align}
    0\otimes1=1,
    \qquad
    C^{10}_{00,10}\neq0.
\end{align}
Thus, the irreducible decomposition of $\qty(H_{\mathrm{mag}})^3$ contains a rank-1, $Q=0$ component.
The spin selection rule therefore allows $M_0^{(3)}$.

These results give the $\mathrm{SU}(2)$ selection rules summarized in Tables~\ref{tab:single_cluster_oh_selection_rules} and \ref{tab:single_cluster_d3d_selection_rules}.

\twocolumngrid
\bibliography{nips3_optical_activation_references.bib}

\end{document}